\documentclass[epsfig,12pt]{article}
\usepackage{epsfig}
\usepackage{graphicx}
\usepackage{hyperref} 
\usepackage{multirow}
\graphicspath{{pics/}}
\DeclareGraphicsExtensions{.pdf,.png,.jpg}

\usepackage{float} 

\usepackage[utf8]{inputenc}
\usepackage[T2A]{fontenc}
\usepackage[english]{babel}

\usepackage{csquotes}

\usepackage{extarrows}
\usepackage{youngtab} 

\usepackage{array}
\usepackage{amsmath}
\usepackage{amssymb}
\numberwithin{equation}{section}

\usepackage{mathtools}
\mathtoolsset{showonlyrefs} 
\usepackage{dynkin-diagrams} 
\newcommand{\invol}[2]{\draw[latex-latex] (root #1) to [out=60,in=120] 
	node[midway,below]{} (root #2);} 

\newcommand{\beq}{\begin{equation}}   
	\newcommand{\eeq}{\end{equation}}
\newcommand{\beqn}{\begin{eqnarray}}   
	\newcommand{\eeqn}{\end{eqnarray}}

\usepackage{xcolor}

\begin{document}
	
	\hypersetup{%
		linkbordercolor=blue,
	}
	
	\unitlength = 1mm
	
	\def\de{\partial}
	\def\Tr{ \hbox{\rm Tr}}
	\def\const{\hbox {\rm const.}}  
	\def\o{\over}
	\def\im{\hbox{\rm Im}}
	\def\re{\hbox{\rm Re}}
	\def\bra{\langle}\def\ket{\rangle}
	\def\Arg{\hbox {\rm Arg}}
	\def\Re{\hbox {\rm Re}}
	\def\Im{\hbox {\rm Im}}
	\def\diag{\hbox{\rm diag}}

	
	\def\QATOPD#1#2#3#4{{#3 \atopwithdelims#1#2 #4}}
	\def\stackunder#1#2{\mathrel{\mathop{#2}\limits_{#1}}}
	\def\stackreb#1#2{\mathrel{\mathop{#2}\limits_{#1}}}
	\def\Tr{{\rm Tr}}
	\def\res{{\rm res}}
	\def\Bf#1{\mbox{\boldmath $#1$}}
	\def\balpha{{\Bf\alpha}}
	\def\bbeta{{\Bf\beta}}
	\def\bgamma{{\Bf\gamma}}
	\def\bnu{{\Bf\nu}}
	\def\bmu{{\Bf\mu}}
	\def\bphi{{\Bf\phi}}
	\def\bPhi{{\Bf\Phi}}
	\def\bomega{{\Bf\omega}}
	\def\blambda{{\Bf\lambda}}
	\def\brho{{\Bf\rho}}
	\def\bsigma{{\bfit\sigma}}
	\def\bxi{{\Bf\xi}}
	\def\bbeta{{\Bf\eta}}
	\def\d{\partial}
	\def\der#1#2{\frac{\d{#1}}{\d{#2}}}
	\def\Im{{\rm Im}}
	\def\Re{{\rm Re}}
	\def\rank{{\rm rank}}
	\def\diag{{\rm diag}}
	\def\2{{1\over 2}}
	\def\ntwo{${\mathcal N}=2\;$}
	\def\nfour{${\mathcal N}=4\;$}
	\def\none{${\mathcal N}=1\;$}
	\def\ntwot{${\mathcal N}=(2,2)\;$}
	\def\ntwoo{${\mathcal N}=(0,2)\;$}
	\def\x{\stackrel{\otimes}{,}}

	\newcommand{\cpn}{CP$(N-1)\;$}
	\newcommand{\wcpn}{wCP$_{N,\widetilde{N}}(N_f-1)\;$}
	\newcommand{\wcpd}{wCP$_{\widetilde{N},N}(N_f-1)\;$}
	\newcommand{\wcpN}{$\mathbb{WCP}(N,N)\;$}
	\newcommand{\wcpK}{$\mathbb{WCP}(K,K)\;$}
	\newcommand{\wcpt}{$\mathbb{WCP}(2,2)\;$}
	\newcommand{\wcpf}{$\mathbb{WCP}(4,4)\;$}
	\newcommand{\wcpo}{$\mathbb{WCP}(1,1)\;$}
	\newcommand{\wcp}{$\mathbb{WCP}(N,\tilde N)\;$}
	\newcommand{\vp}{\varphi}
	\newcommand{\pt}{\partial}
	\newcommand{\tN}{\widetilde{N}}
	\newcommand{\ve}{\varepsilon}

	\newcommand{\sun}{SU$(N)\;$}
	
	\newcommand\xdownarrow[1][2ex]{%
		\mathrel{\rotatebox{90}{$\xleftarrow{\rule{#1}{0pt}}$}}
	}
	\newcommand\xupdownarrow[1][2ex]{%
		\mathrel{\rotatebox{90}{$\xleftrightarrow{\rule{#1}{0pt}}$}}
	}
	

	\begin{titlepage}

		\vspace*{-3cm}
		\begin{flushright}
			{\footnotesize 
				FTPI-MINN-26-11 \\ UMN-TH-4531/26 
			}
		\end{flushright} 
		
		\vspace{1cm}
		
		\begin{center}
			{  \Large \bf  
				Hadrons in $\mathcal{N}=2$ supersymmetric QCD \\[5pt] from non-Abelian string on  2D black hole 
			}

			
			\vspace{5mm}
			
			{\large  \bf E.~Ievlev,$^{\,a}$ A.~Marshakov,$^{\,\,b,c}$ G.~Sumbatian,$^{\,d}$\\  and  A.~Yung$^{\,d,e}$}
		\end{center}

		\begin{center}
			
			{\it  $^{a}$William I. Fine Theoretical Physics Institute,
				University of Minnesota,
				Minneapolis, Minnesota 55455, USA}\\
			
			$^{b}${\it Skoltech,
				Moscow 121205, Russia
			}\\
			$^{c}${\it Dept. Math., HSE University,
				Moscow 119048, Russia}
			
			$^{d}${\it NRC ``Kurchatov Institute'' - PNPI,  Gatchina, St. Petersburg
				188300, Russia}\\
			
			{\it $^{e} $HSE University, St. Petersburg,
				194100, Russia}
			
		\end{center}

		
\begin{center}
	{\large\bf Abstract}
\end{center}
\noindent
We continue the study of non-Abelian vortex string in 4D $\mathcal{N}=2$ supersymmetric QCD as critical superstring, and extend this analysis to $U(N)$ gauge theory with arbitrary even $N$ and $N_f=2N$ number of quarks. We  introduce a special mass deformation and show that the SQCD hadron spectrum is still given by the string spectrum on the 2D $\mathcal{N}=2$ supersymmetric black hole. We perform a cross-check by computing the multiplicity of hadronic states of the  high-energy part of the spectrum  both from string and field theory pictures. We also clarify the spontaneous breaking of the global flavor symmetry by VEV of the massless baryon. We finally claim, that phase diagram of $\mathcal{N}=2$ SQCD with $N_f=2N$ consists of the Higgs phase at weak coupling and string/hadronic phase at strong coupling, separated by phase transition, and is seen as a conifold transition from string theory point of view.

	\end{titlepage}
	
	\newpage
	
	\tableofcontents

	\newpage
	\section{Introduction}
	\label{sec:intro}

	Non-Abelian vortex strings in four-dimensional 
	(4D) $\mathcal{N}=2$  supersymmetric QCD (SQCD) \cite{HT1,ABEKY,SYmon,HT2}, are flux tube configurations, saturating BPS bound, see \cite{Trev,Jrev,SYrev,Trev2} for reviews.
	Besides usual translational moduli, they also have  (internal) orientational and size zero modes, comprising a two-dimensional (2D) world sheet sigma model, the so-called \ntwot supersymmetric weighted $\mathbb{CP}$-model $\mathbb{WCP}(N,N_f-N)$.

	It was shown in \cite{SYcstring} that non-Abelian solitonic string in \ntwo SQCD with gauge group U(2) and $N_f=2N=4$ quark flavors behaves as a critical superstring\footnote{By “critical superstring” we mean a canonical superstring with \none local supersymmetry in the world-sheet (super)gravity sector; the \ntwot supersymmetric world-sheet models (including \ntwo Liouville) belong to the matter sector.}. For $N_f=2N$ the world-sheet $\mathbb{WCP}(N,N)$ sigma model is conformal, and for $N=2$ the six orientational/size moduli together with four translational moduli form a ten-dimensional target space $\mathbb{R}^4\times Y_6$ required for a superstring to become
	critical \cite{SYcstring,KSYconifold}. The target geometry contains the conifold $Y_6$ -- the noncompact Calabi-Yau (CY) space  \cite{Candel,NVafa}. The resulting critical vortex string is type IIA, and its low-lying closed-string spectrum was found in \cite{KSYconifold,SYlittles}.
	
	Most string modes are non-normalizable over $Y_6$, so they are not localized in 4D; in particular, there is no 4D graviton. An exception is the conifold complex-structure deformation modulus $b$, whose wave function is logarithmically normalizable and was interpreted as a massless baryon in 4D \ntwo SQCD \cite{KSYconifold}. 
	
	Massive states were also analyzed using \ntwo Liouville theory \cite{Ivanov,KutSeib}, by an approach akin to that in little string theories \cite{Kutasov}, based on the conjectured equivalence \cite{GVafa,GivKut,GivKutP} between the critical string on conifold and noncritical $c=1$ string with compact scalar on the self-dual radius plus a Liouville field.
	This equivalence was used in \cite{SYlittles,SYlittmult} to obtain the low-lying hadron spectrum of 4D \ntwo SQCD with U(2) and $N_f=4$.
	
	More recently, in \cite{GIMMY} a direct derivation was presented: a Coulomb branch opening up at strong coupling in $\mathbb{WCP}(N,N)$ model on noncompact toric CY manifolds is described by \ntwo Liouville theory with $N$-dependent background charge. Using this Liouville description, in \cite{Y_mass_Liouville} an interpolation program between SQCDs with different gauge groups/quark flavor numbers by turning on quark masses and decoupling (or integrating in) quark flavors was initiated. In particular, one can connect in this way the U(2) theory with $N_f=4$ with U(4) and $N_f=8$ SQCD by lowering the masses of four extra flavors.
	
	Quark masses break the world-sheet conformal invariance, so the mass-deformed theory cannot be quantized directly; instead one finds true string vacuum by solving the effective supergravity equations (vanishing of certain beta functions in world-sheet theory) with initial conditions set by the mass deformation \cite{Y_mass_Liouville}. Since the relation $N_f=2N$ is kept intact by our mass deformation, the coupling constant in the \wcpN model does not run, and the world-sheet conformal invariance is then ensured by finding  the effective  gravity solution.
	
	It was shown in \cite{IMSY} that the mass-deformed background found in  \cite{Y_mass_Liouville} has actually the geometry of a trumpet, first discussed in \cite{Giveon,DijVerVer}. It is $T$-dual to the 2D \ntwo  black hole with cigar metric, described exactly by \ntwo coset SL$(2,\mathbb{R})$/U(1) model \cite{GVafa,GivKut,Wbh,MukVafa,OoguriVafa95}. Since \ntwo Liouville theory is mirror to the \ntwo cigar \cite{HoriKapustin}, it was  argued in \cite{IMSY} that the theory with both superpotential and mass deformations included still yields the same cigar geometry, and only the black-hole mass or entropy gets now contributions from both deformations, being dominated by each of them in corresponding regime.
	
	This framework allowed us to extract the  hadron spectrum of \ntwo SQCD with gauge group U(4) and $N_f=8$. Remarkably, the functional form of the string spectrum turned out to be insensitive to the mass deformation; only the multiplicities of stringy states on each energy level change, which was shown to be  qualitatively consistent with SQCD-side field-theory arguments \cite{IMSY}. 
	
	In this paper we extend this line of research and generalize this construction to the case with an arbitrary even number of colors $N$.
	In other words, we take \ntwo SQCD with gauge group $U(N)$ and $N_f=2N$ and use a special interpolation procedure, generalizing that from \cite{Y_mass_Liouville,IMSY}. 
	We analyze the hadron spectrum both from the field theory side and from the string theory perspective and find excellent agreement.
	This serves as another strong check of our description of the non-Abelian flux tube as a critical superstring.
	
	First we find that the SQCD hadron spectrum does not depend on $N$ and is still given by the string spectrum of the cigar or SL$(2,\mathbb{R})$/U(1) coset with the same Ka\v{c}-Moody level $k=1$. Then we make a quantitative check
	of the multiplicities of states  in large $N$ limit both from string theory and SQCD  sides and find an exact agreement.  From 
	string theory side the growth of number of hadronic states with $N$ is found from  the  near-Hagedorn behavior of 2D black hole.
	
	We also check that stringy massless $b$-baryons  can be viewed as Goldstone states, present due to spontaneous breaking of the global symmetry  by vacuum expectation value (VEV)  of the $b$-baryon.   This check also confirms the self-consistency of our approach. 
	
\begin{figure}[h]
		\centering
		\includegraphics[width=0.7\linewidth]{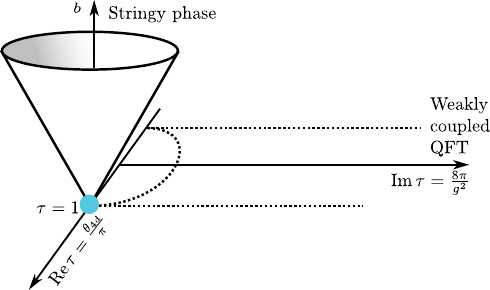}
		\caption{
			The phase diagram for 4D SQCD. The fundamental domain of 4D coupling $\tau_{SW}$ is shown in the horizontal plane. SQCD is in the Higgs phase on this plane. The stringy phase where $\langle b\rangle \neq 0$ is  schematically  shown by the cone.
		}
		\label{fig:phase_diagram}
	\end{figure}

	Qualitatively the emergent picture can be summarized by  \ntwo SQCD phase diagram shown in Fig.~\ref{fig:phase_diagram}. Physics depends on the value of the complexified (by four dimensional theta-angle $\theta_{4D}$) non-Abelian gauge coupling constant \cite{SW2}
	\beq
	\tau_{SW}= i\frac{8\pi}{g^2} + \frac{\theta_{4D}}{\pi},
	\label{tau_SW}
	\eeq
	which does not run in \ntwo SQCD with $N_f=2N$. 
	
	At small $g^2$ (large $\text{Im}\tau_{SW}$)  one has a perturbative Higgs phase, where the physical spectrum consists of a finite number of screened quarks and Higgsed gauge bosons together with their superpartners,  see Sec. \ref{sec:SQCD} and \cite{SYrev} for details. 
	
	The  region of $ g^2 \sim 1$ is at strong coupling. In particular, at the ''super strong''
	point   $\tau_{SW} =1$ a non-perturbative Higgs branch is developed as schematically shown in Fig.~\ref{fig:phase_diagram}. On this non-perturbative branch the massless stringy baryon $b$, associated with the complex structure modulus of the conifold, develops a  VEV. Here we have a stringy phase with towers of massive string states, see Sec. \ref{sec:hadrons_from_primary}. 
	
	Many years ago it was proposed in \cite{FradShen}, that 
	there is no phase transition between Higgs phase at weak coupling and confinement phase at strong coupling in scalar QCD~\footnote{The reason (motivated by lattice studies)
		is that there is no order parameter to distinguish  between these phases, since Wilson loop in a theory with dynamical quarks in the fundamental representation of gauge group has a perimeter law in both phases, due to the quark pair creation which breaks the confining string.}.
	Motivated by dynamics  of 4D \ntwo SQCD, we challenge this conclusion. In our theory the hadron phase at strong coupling can be described  by string theory of non-Abelian vortices, responsible for the confinement of monopoles. We 
	show that although both phases do have massless states, these states belong to different representations of the global symmetry group, and therefore these two phases are not analytically connected. They become separated by a phase transition, seen from the string theory point of view as famous conifold transition.

	It is also worth noting that there is a 
	qualitative distinction between the two phases. 
	As we mentioned above, in  the Higgs phase at weak coupling the physical spectrum is formed by a finite
	number of screened quarks and Higgsed gluons together with their superpartners. In contrast, the stringy phase has an infinite tower of
	4D hadron  states,  typical for the spectrum in string theory. The higher-energy states of this spectrum cannot be interpreted as composites  of a finite number of light states.

	One can even speculate, that in real-world QCD the hadronic phase is in the same universality class as the stringy phase of \ntwo SQCD.  Historically, the difficulties of describing Regge behavior of hadron trajectories by field theory methods led to the creation of string theory.  Here the non-perturbative hadron spectrum of \ntwo SQCD in stringy phase is determined and computed using the world-sheet theory of the critical superstring.

	The paper is organized as follows. In review Sec.~\ref{sec:NAstring} we first introduce 4D \ntwo SQCD and the world sheet \wcpN model on the non-Abelian string. Next we 
	discuss \ntwo Liouville theory  describing the Coulomb branch of \wcpt model, which opens up at strong coupling. We also review the string spectrum of 2D \ntwo black hole, mirror to \ntwo Liouville theory. In Sec.~\ref{Sec:Mass_def} we study the special interpolation procedure:  consider the mass-deformed background with trumpet metric and its dual description in terms of the 2D black hole. We also discuss the resulting phase diagram of 4D \ntwo SQCD with $N_f=2N$. In Sec.~\ref{sec:omega} we consider the multiplicity of hadronic states from both string theory and SQCD sides in the large $N$ limit. In Sec.~\ref{Sec:Goldstones} we show that massless $b$-baryons can be viewed as Goldstone bosons upon global flavor symmetry breaking by the VEV of $b$-baryon. Sec.~\ref{sec:concl} contains our conclusions and appendices present some details of our calculations.

	\section {Non-Abelian vortex string and $\mathcal{N}=2$ Liouville theory}
	\label{sec:NAstring}

	In this Section we introduce the 4D SQCD with a specific mass deformation.
	Then we review some basic facts about the world sheet theory of non-Abelian vortex string and the hadron spectrum in this context, before plunging into concrete computations in the rest of this paper.

	\subsection{Four-dimensional \boldmath{${\mathcal N}=2$} SQCD}
	\label{sec:SQCD}
	
	Our main theory of interest is the $\mathcal{N}=2$ SQCD in 4D with eight supercharges; see, for example, \cite{SYrev} for a detailed review of this theory.
	We focus on the case with the gauge group $U(N)$ and $N_f = 2 N$ matter hypermultiplets in the fundamental representation,  with this choice the 4D gauge coupling does not run.  However, the conformal invariance of the 4D SQCD is explicitly broken by quark masses and
	the Fayet-Iliopoulos (FI) term  \cite{FI}, with FI parameter $\xi$, which is not renormalized and defines  VEVs of quarks.

	At weak coupling $g^2\ll 1$, this theory is in the Higgs phase. 
	In a vacuum where the first $N$ quark flavors are massless at zero $\xi$, the matrix of adjoint scalars of the $\mathcal{N}=2$ vector multiplet develops VEV of the form
	\begin{equation}
		\langle a \rangle = -
		\text{diag} (m_1 \,, \ldots \,, m_N) \,,
		\label{avev}
	\end{equation}
	where  $m_A$ ($A=1,..., N_f$) are bare quark masses. Adjoint condensates \eqref{avev} break U$(N)$ gauge group down to U(1)$^N$, with the masses of the off-diagonal gauge bosons given by 
	$|m_k-m_l|$ ($k,l=1,..., N$), while the quark masses of $q^{kA}$ and $\tilde{q}_{Ak}$ (two complex scalars of the $\mathcal{N}=2$ hypermultiplet) are equal to $|m_k-m_A|$. 
	
	At nonzero $\xi$, first $N$ squarks also develop the VEV's
	\begin{equation}
		\langle q^{kA}\rangle =\sqrt{
			\xi}\, \delta^{kA}, \qquad \langle \tilde{q}_{Ak} \rangle =0\qquad k=1,...,N ,\; A=1,...,N_f.
		\label{qvev}
	\end{equation}
	%
	%
	%
	These  condensates break the U$(N)$ gauge group, Higgsing  all gauge bosons. 
	The quarks are screened, while the monopoles are confined by non-Abelian strings~\footnote{In fact, in  U$(N)$ theories, confined monopoles are junctions of two distinct elementary non-Abelian strings, see \cite{SYrev} for a review. In particular, the baryons represent closed \textquote{necklace}
		configurations of monopoles on closed string.}. 
	At weak coupling, these stringy hadrons are heavy and decay into perturbative states; however, at strong coupling, the theory enters the so-called instead-of-confinement regime \cite{SYi_of_c,ISY_b_baryon}. In this regime quarks and gluons decay into monopole-antimonopole pairs and we are left with hadrons formed by monopoles confined by non-Abelian strings.
	
	Below we concentrate on the theories with arbitrary even $N$, $N=2K$ with some integer $K$.
	We also consider a particular $\mathbb{Z}_K$-symmetric set of bare quark masses,
	\begin{equation}
		\{ m_A \}_{A=1}^{N_f} =\{\underbrace{m_2,m_2,\ldots,m_{2K},m_{2K;}}_{N=2K}\underbrace{\tilde{m}_2,\tilde{m}_2,\ldots,\tilde{m}_{2K},\tilde{m}_{2K}}_{N=2K}\}, \quad 
		\tilde{m}_{2k}=m_{2k} ,
		\label{mass_choice}
	\end{equation}
	%
	%
	where we take $m_{2k-1}=m_{2k}$, $k=1,...,K$
	with the parameters $m_{2k}$ evenly spaced on a circle of radius $m\in \mathbb{R}_{\geq 0}$,
	\begin{equation}
		m_{2k}=m \exp\left(\frac{2\pi i k}{K}\right),\quad k=1,\ldots K
		\label{Zp_mass}
	\end{equation} 
	This particular mass choice will be motivated in Sec.~\ref{Sec:Mass_def_Liouv} below.

	At large $m$, the original SQCD splits into $K$ sectors with mutual interactions suppressed by the energy scale $\Delta m \sim m/K$,
	\begin{equation}
		[U(2K) + (N_f=4K)] \xrightarrow{ m \to \infty } [U(2) + (N_f=4)]^{\otimes K}
		\label{M_to_infty}
	\end{equation}
	(see \cite{Y_mass_Liouville} for details).  Below in this  paper we often consider the large $K$ limit, still we assume that $m/K \to \infty$  at the initial step of our mass interpolation procedure.

	For $m \to 0$ we recover the full SU$(N=2K)$ gauge group with $N_f=4K$ fundamental flavors.
	In this case, the global flavor SU$(N_f)$ is also broken down by quark VEV's to the so-called color-flavor
	locked group. The resulting global symmetry is
	\beq
	SU(N)_{C+F}\times SU(N)\times U(1)_B,
	\label{c+f}
	\eeq
	see \cite{SYrev} for more details. 
	The unbroken global U(1)$_B$ factor above is identified with a baryonic symmetry.

	\subsection{World-sheet theory of the non-Abelian  string}
	\label{sec:wcp}
	
	In \ntwo SQCD the flux-tube strings are 1/2-BPS saturated and preserve \ntwot supersymmetry with four supercharges on the world sheet.
	Their tension is determined by the 4D FI parameter
	\beq
	\tau=2\pi \xi = \frac1{2\pi\alpha'}
	\label{ten}
	\eeq 
	and in addition to translational moduli the internal modes form an extra world-sheet sigma-model,
	which we review below.
	
	\subsubsection{\wcpN model}
	
	The effective theory on the string world sheet for the internal moduli is 2D \ntwot supersymmetric \wcpN model, defined  as a low-energy limit of the  U(1) gauge theory \cite{W93}
	with twisted-mass deformation
	\begin{equation}
		\begin{aligned}
			&S = \int d^2 x \left\{
			\left|\nabla_{\alpha} n^{i}\right|^2 
			+\left|\widetilde{\nabla}_{\alpha} \rho^j\right|^2
			-\frac1{4e_0^2}F^2_{\alpha\beta} + \frac1{e_0^2}\,
			\left|\pt_{\alpha}\sigma\right|^2 + \frac1{2e_0^2}\,D^2- \frac{\Theta}{2\pi}F_{01}
			\right.
			\\[3mm]
			&-
			\left.\left|\sqrt{2}\sigma +m_i\right|^2 \left|n^{i}\right|^2 - \left|\sqrt{2}\sigma +\tilde{m}_j\right|^2 \left|\rho^j\right|^2
			+D\left(\left|n^{i}\right|^2-\left|\rho^j\right|^2 - {\rm Re}\,\beta \right)\right\},
			\\[3mm]
			&  \alpha,\beta=0,1\,,\quad i,j=1,...,N, 
		\end{aligned}
		\label{wcpNN}
	\end{equation}
	see review \cite{SYrev} for details. 
	Here, the complex fields $n^{i}$ and $\rho^j$ describe the orientational and size moduli respectively \cite{HT1,HT2,AchVas,SYsem,Jsem,SVY}. 
	They have charges  $+1$ and $-1$, so that
	\begin{equation}
		\nabla_{\alpha}=\pt_{\alpha}-iA_{\alpha}\,,
		\qquad 
		\widetilde{\nabla}_{\alpha}=\pt_{\alpha}+iA_{\alpha}\,.	
		\label{cov_derivatives}
	\end{equation}
	Twisted masses of the $n^i$ and $\rho^j$ fields coincide with the masses $m_i$ and $\tilde{m_j}$ of the  4D quarks. 
	
	The complexified inverse coupling in \eqref{wcpNN},   
	\begin{equation}
		\beta = {\rm Re}\,\beta + i \, \frac{\Theta}{2 \pi} \,
		\label{beta_complexified}	
	\end{equation}
	is defined via the  2D FI term  (twisted superpotential)
	\beq
	\frac{\beta}{2}\,\int d^2 \tilde{\theta}\sqrt{2}\,\Sigma = \frac{\beta}{2}\,(D-iF_{01})
	\label{Sigma_sup},
	\eeq
	where $\Theta$ is the 2D $\theta$-angle, while $\Sigma$ is the twisted chiral superfield for the 2D gauge multiplet which contains complex scalar $\sigma$, gauge field strength and
	auxiliary field $D$ as components \cite{W93}.
	
	The global symmetry group of the \wcpN sigma model coincides with the unbroken global group of the 4D SQCD \eqref{c+f}; see \cite{KSYconifold}.
	The fields $n^i$ and
	$\rho^j$ transform in   the representations 
	\beq
	\left(\,{\bf N}, \,{\bf 1}, \,\frac12 \right) \qquad \left(\,{\bf 1},\,{\bf N} , \,\frac12 \right)
	\label{n_rho_representations}
	\eeq
	respectively.
	
	The BPS spectrum of states in this 2D world sheet theory coincides with the BPS spectrum of 4D states in the quark vacuum \eqref{qvev} given by the exact Seiberg-Witten solution \cite{SW2} at $\xi=0$. This coincidence was observed in \cite{Dorey,DoHoTo} and explained
	later in \cite{SYmon,HT2} using the picture of 4D monopoles confined by the non-Abelian string that are seen as kinks in the 2D world-sheet theory.  
	It is called 2D-4D correspondence. 
	
	The beta function vanishes in the \wcpN model and  this theory is conformal in the massless limit. Therefore, its target space is Ricci-flat and K\"ahler, due to \ntwot supersymmetry, i.e. it is  a (noncompact) Calabi-Yau  (CY) manifold.
	
	Let us discuss how the interpolation procedure (discussed around Eq.~\eqref{M_to_infty}) looks like from the world-sheet point of view. Consider  classical vacua of the \wcpN model. At $\mathrm{Re}\,\beta>0$ we have $N$ vacua of the world sheet theory, parametrized by the VEV of $\sigma$,
	\begin{equation}
		\sqrt{2}\sigma=-m_{i_0},\quad |n^{i_0}|^2={\rm Re}\,\beta, \quad i_0=1,\ldots,N.
		\label{class_vacua}
	\end{equation}
where each choice of $i_0$ corresponds to a particular vacuum of the 2D worldsheet theory.
	Masses of the perturbative excitations $n^i,\, i\neq i_0$ and $\rho^j$ are given by exact BPS  formulas $|m_i-m_{i_0}|$ and $|\tilde{m}_j-m_{i_0}|$   respectively. The number of vacua remains intact in quantum theory, since it is protected by Witten index, equal to $N$.
	 For the particular set of masses \eqref{mass_choice}, all  $n^i$ and $\rho^i$ fields  (except for $i = i_0$ and one of the neighbours $i= i_0\pm 1$) decouple in the $m/K\to\infty$ limit, and the $\mathbb{WCP}(N,N)$ model \eqref{wcpNN} reduces to \wcpt being the starting point of the interpolation procedure.
	
	The target space of the \wcpt is a non-compact CY$_3$ manifold, namely the conifold \cite{KSYconifold,NVafa}. Combined with the flat Minkowski 4D space, it  forms a ten-dimensional  space required for a superstring to be critical \cite{SYcstring,KSYconifold}.  
 The conifold $Y_6$ is a cone with a section $S^2\times S^3$ at a fixed radial coordinate. The conifold singularity   can be smoothed out
 		in two distinct ways: by resolving the singularity with \(\mathbb{CP}^1\) with a non-vanishing K\"ahler class
 		or by  deforming the 
 		complex structure; see  \cite{Candel,NVafa}. The first option, called the resolved conifold, amounts to keeping
 		a nonzero value of $\beta$ in the $D$-term condition 
 		\beq
 		|n^{i}|^2-|\rho^j|^2 = {\rm Re}\,\beta,
 		\label{D_term_condition}
 		\eeq
 		see  \eqref{wcpNN} for $N=2$. 
 		For $\rho^j=0$,   this gives exactly the $\mathbb{CP}^1$ part of the target space, and  $\Re\beta$ defines the size of the ``blown up'' $\mathbb{CP}^1\simeq S^2$.  
 		The resolved conifold has no normalizable zero modes. 
 		In particular, 
 		the K\"ahler modulus $\beta$,  which, in principle, could become a massless scalar field in four dimensions, has a non-normalizable (quadratically divergent) wave function over the 
 		$Y_6$ and therefore is not dynamical \cite{KSYconifold}. 

	However, we are interested in considering the model at ``super strong'' coupling $\beta=0$, where the so-called ``thin string conjecture'' is applicable \cite{SYcstring,KSYconifold}, which implies that at vanishing $\beta$  the world sheet model can be treated as consistent fundamental string theory. 
	
	The conifold singularity ${\rm det}\, w^{ij} =0$, where $w^{ij} =n^i \rho^j$ is the set of U(1) gauge-invariant ``mesonic'' variables, can be alternatively smoothed at $\beta=0$ by the deformation of complex structure. Following  \cite{Candel,NVafa}, this deformation can be described as 
	\beq
	{\rm det}\, w^{ij} =b,
	\label{deformedconi}
	\eeq
	where $b$ is a complex parameter of  deformed conifold. 
	Promoted to a 4D field, parameter  $b$  was interpreted as a scalar component of a  massless baryonic hypermultiplet  of 4D \ntwo QCD in  \cite{KSYconifold}. Its quantum numbers with respect to the global symmetry group \eqref{c+f} for  $N=2$ are
	determined by \eqref{deformedconi},
	\beq
	\left(\,{\bf 1}, \,{\bf 1}, \, 2 \right),
	\label{b_representation}
	\eeq 
	i.e. it is a singlet with respect to both $SU(2)$ groups with $B(b)=2$ baryonic  charge \cite{KSYconifold}.
	The massless field $b$  can form a condensate. Thus, 
	we have a new Higgs branch in 4D \ntwo SQCD which develops only at the critical value of 
	the 4D coupling constant $\tau_{SW}=1$ associated with $\beta=0$ \cite{IYcorrelators}, see Fig.1 and discussion in the Introduction.

	\subsubsection{\ntwo Liouville theory}

	It was recently shown \cite{GIMMY}, that the Coulomb branch of 2D \wcpN world sheet model which opens up at the  strong coupling point $\beta=0$ can be described by \ntwo Liouville theory (see \cite{Nakayama} for a review of Liouville theory in general).
	Its bosonic action reads
	\beq
	S_{\rm eff}=\frac{1}{4\pi}\int d^2 x\sqrt{h}
	\;\left(\frac1{2}\,h^{\alpha\beta}(\pt_{\alpha}\phi\pt_{\beta}\phi  +\pt_{\alpha}Y\pt_{\beta}Y)
	-\frac{Q}{2}\phi\, R^{(2)} \right),
	\label{Liouville}
	\eeq
	for the real Liouville scalar field $\phi$, supplemented by the real compact scalar $Y \sim Y+2\pi$. Here $R^{(2)}$ is the Ricci scalar for the world sheet metric
	$h_{\alpha\beta}$,  and  $h={\rm det} (h_{\alpha\beta})$.  
	The linear dilaton term 
	\beq
	\Phi =-\frac{Q}2\,\phi
	\label{linear_dilaton}
	\eeq
	contains the background charge $Q=\sqrt{2(N-1)}$ , so for the conifold $N=2$ case $Q=\sqrt{2}$. The  \ntwo Liouville central charge is
	$c_L=3(1+Q^2)$, so that in $N=2$ case $\left.c_L\right|_{N=2}=9$ as required for criticality.

	The \ntwo Liouville interaction superpotential (a marginal deformation of the \ntwo Liouville theory \eqref{Liouville}, see \cite{Nakayama}) comes from  the 2D FI term \eqref{Sigma_sup} in the \wcpt model and
	has the form 
	\beq
	L_{int}= b \,\int d^2 \tilde{\theta}\,e^{-\frac{\phi +iY}{Q}}
	\label{Liouville_sup},
	\eeq
	where the scalars $\phi$ and $Y$ are promoted to (twisted) chiral superfields, and $b$ is the conifold complex structure deformation parameter \eqref{deformedconi}. To get \eqref{Liouville_sup} we use the relation
	\beq
	\sigma=\gamma\, e^{-\frac{\phi + iY}{Q}},
	\label{sigma}
	\eeq
	between the complex scalar $\sigma$ from the U(1) gauge multiplet and the fields $\phi$ and $Y$ in the Liouville theory, where $\gamma = -\sqrt{2}\,b/\beta$. Note that constant $\gamma$ is singular at $\beta\to 0$ in order to keep the Liouville superpotential \eqref{Liouville_sup} finite, see \cite{GIMMY}
	for details.
	
	The string coupling constant $g_s=e^{\Phi}$, determined by linear dilaton \eqref{linear_dilaton}, grows at large negative $\phi$. On the other hand  at nonvanishing $b$, the Liouville wall 
	prevents the field $\phi$ from penetrating into the region of large negative values. For $Q=\sqrt{2}$ the string coupling 
	$g_s\sim 1/|b|$, and below we keep $b$
	large to ensure that the string
	perturbation theory is reliable, see \cite{GivKut,SYlittmult}.
	In terms of 4D SQCD, at large $b$ 
	we are on the non-perturbative  Higgs branch far away from the origin, see SQCD phase diagram in Fig. 1.

	\subsection{SQCD hadrons from Liouville theory}
	\label{sec:hadrons_from_primary}
	
	The primary operators of the \ntwo Liouville theory for $N=2$ ($Q = \sqrt{2}$) case describe physical string states, interpreted as  hadrons in 4D SQCD; see \cite{SYlittles} for details.
	The operators relevant for our 4D applications have the form
	\cite{GivKut,Teschner:1999ug,LSZinLST}
	\begin{equation}
		T_{j, m}
		\simeq
		e^{i Qm Y } e^{Q j \phi}
		\label{genericj}
	\end{equation}
	with the conformal dimensions
	\beq
	\Delta_{j,m} = \frac{Q^2}{2}\left\{m^2 - j(j+1)\right\} = \frac1{k}\left\{m^2 - j(j+1)\right\}\, ,
	\label{dimV}
	\eeq
	The spectrum of allowed values of  $j$ and $m$  in \eqref{genericj} is exactly determined, using the
	mirror description  \cite{HoriKapustin} of the theory as a  ${\rm SL}(2,R)/U(1)$ coset 
	with the level
	\beq
	k=\frac{2}{Q^2}
	\label{k_Q}
	\eeq
	of the \ntwo supersymmetric version of the Ka\v{c}-Moody algebra
	in \cite{MukVafa,DixonPeskinLy,Petrop,Hwang,EGPerry}, 
	see \cite{EGPerry-rev} for a review.
	
	We are interested in  the  discrete spectrum
	\begin{equation}
		j=-\frac12, -1; \qquad m=\pm\{j, j-1,j-2,...\},
		\label{discrete}
	\end{equation}
	where the restriction to two values of $j$ comes from unitarity bound for $\left.k\right|_{Q = \sqrt{2}}=1$ \cite{SYlittles}.
	
\begin{figure}
		\epsfxsize=6cm
		\centerline{\epsfbox{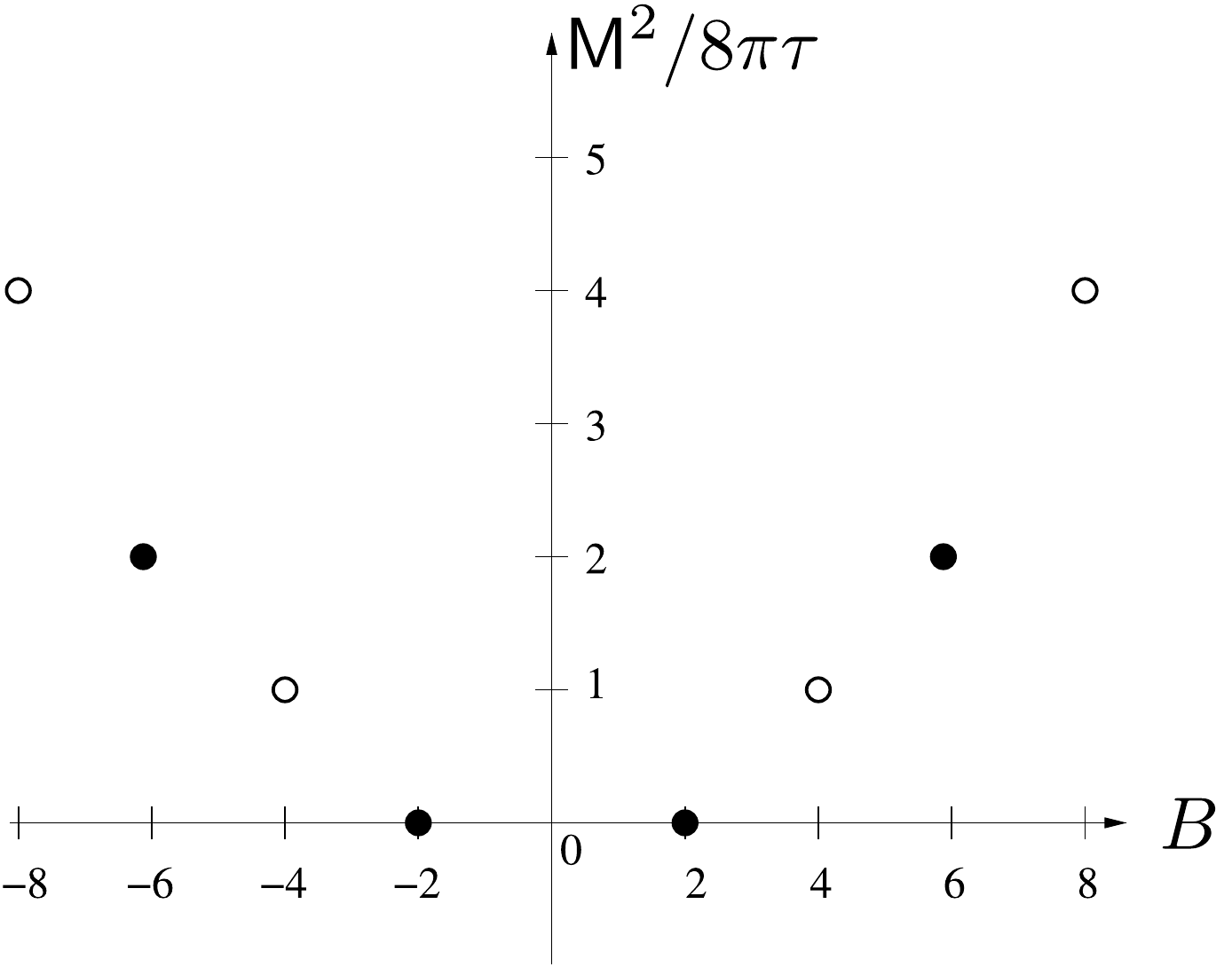}}
		\caption{\small  Spectrum of spin-0 and spin-2 states \eqref{tachgrmass} as function of the baryonic charge \eqref{m-baryon}. Closed 
			and open circles denote  spin-0 and spin-2 states, respectively.
		}
		\label{fig_spectrum}
	\end{figure}

	Dressing the  vertex \eqref{genericj} with the 4D plane wave, one obtains the operators corresponding to 4D states.
	For the scalar and spin-2 states we have 
	\beq
	\mathcal{T}_{j,m}=e^{ip_{\mu}x^{\mu}}\, T_{j,m} 
	\quad  \text{ and } \quad
	V_{j,m}^{\mu\nu}=  \psi^{\mu}_L\psi^{\nu}_R\, 
	e^{ip_{\mu}x^{\mu}}\, T_{j;m}, 
	\label{dressed_tachyon}
	\eeq
	where $ \psi^{\mu}_{L,R}$ are world-sheet  fermionic superpartners of $x^{\mu}$.
	Imposing the mass-shell condition 
	\begin{equation}
		\mathsf{M}_T^2=-p_{\mu}p^{\mu} = 2\Delta_{j,m}-1 , \qquad \mathsf{M}_V^2 = 2\Delta_{j,m}
		\label{mass_shell}
	\end{equation}
	and the GSO projection \cite{GivKut}, we find the masses (in dimensionless units, imposing normalization $4\pi\tau=1$ or 
	\begin{equation}
		\alpha' =2
		\label{a2}
	\end{equation}
	for the string tension \eqref{ten}) of the 4D scalar and spin-2 states \cite{SYlittles,SYlittmult}:
\begin{align}
		&\mathsf{M}_T^2 = 2 m^2 - \frac12 \,, \quad j=-\frac{1}{2}, \quad m = \pm \frac{1}{2} \,, \pm \frac{3}{2} \,, \ldots
		\\
		&\mathsf{M}_V^2 = 2m^2,  \quad j=-1, \qquad |m|=1,2,... .
		\label{tachgrmass}
\end{align}
	%
	%
	This stringy spectrum of 4D SQCD is shown in Fig. \ref{fig_spectrum} taken from \cite{SYlittles}.
	
	In particular, the scalar vertex in  \eqref{dressed_tachyon} with $j=-1/2$, $m=\pm 1/2$ gives a massless state in strongly coupled 4D SQCD, corresponding  to the complex structure modulus $b$ for the string compactification on the conifold (cf. with \eqref{Liouville_sup}, two values $m=\pm 1/2$ correspond to two real degrees of freedom of the complex scalar). 
	The associated string state has a logarithmically normalizable wave function  in terms of the conifold radial coordinate  \cite{KSYconifold,Strominger_95}. On the Liouville side, this corresponds to the borderline normalization of the massless state \eqref{genericj} with $j=-\frac12$, $m=\pm \frac12$, see \cite{SYlittles} for details.
	
	For these states, $m$ is related to the baryonic charge as
	\cite{SYlittles,SYlittmult}
	\beq
	B = 4m,
	\label{m-baryon}
	\eeq
	and all closed string states are baryons.
	All states in \eqref{dressed_tachyon} are in fact lowest states of the corresponding linear Regge trajectories \cite{SYlittmult}.

	\section{Mass deformation}
	\label{Sec:Mass_def}
	
	In this Section we introduce the mass deformation \eqref{mass_choice} and study its response in  the Liouville world sheet theory.

	\subsection{Effective action and initial conditions}
	\label{Sec:Mass_def_Liouv}
	
	The derivation below follows closely \cite{Y_mass_Liouville}, but generalizes the previous result 
	obtained for $N=4$ to an arbitrary even $N = 2K$.
	We start with the $\mathbb{WCP}(N,N)$ world sheet action \eqref{wcpNN} with non-vanishing mass parameters \eqref{mass_choice}.
	Following \cite{GIMMY} we integrate out the ``matter'' fields $n_i,\rho_i$, which become massive at Coulomb branch of  \wcpN  model where field $\sigma$ can take arbitrary values at $\beta=0$.
	To integrate out $n^i$ and $\rho^j$ we first use the  large $N$  approximation developed by Witten in
	\cite{W79}   
	and take 
	\begin{equation}
		N = N_0 \cdot K
	\end{equation}
	with fixed $K$, while $N_0 \to \infty$ playing the role of large-$N$ parameter, to be set at the end to the critical conifold value $N_0=2$.
	The resulting kinetic part of the one-loop effective action for $\sigma$ field is  \cite{Y_mass_Liouville}
	%
	%
	\begin{equation}
		S_{eff}^\sigma=\frac{1}{8\pi}\int d^2 x \sum_{A=1}^{2N}\frac{|\partial_\alpha \sigma|^2}{\left|\sigma+\frac{m_A}{\sqrt{2}}\right|^2}
		\label{S_eff_gen}
	\end{equation}
	In the vicinity of any vacuum \eqref{class_vacua} (which are all equivalent due to a very
	special choice of $\mathbb{Z}_K$-symmetric  mass deformation \eqref{mass_choice}) let us
	perform a change of variables
	\begin{equation}
		\sigma+\frac{m_{i_0}}{\sqrt{2}}=\gamma_{i_0} e^{-\frac{\phi+iY}{Q}},
		\label{sigma_mod}
	\end{equation}
	c.f. \eqref{sigma}, where $\gamma_{i_0} = -\sqrt{2}\,b_{i_0}/\beta$, while $b_{i_0}$ is the complex structure deformation parameter  of each conifold we start with at $m\to \infty$.
	Evaluating the sum over flavors, we obtain (see Appendix \ref{App:Seff})
	\begin{equation}
		S_{eff}^\sigma=\frac{2N_0}{4\pi Q^2}\int d^2 x g_{cl}(\phi)\left[ \frac{1}{2}(\pt_{\alpha}\phi)^2 + \frac{1}{2}(\pt_{\alpha}Y)^2\right] .
		\label{S_eff}
	\end{equation}
	The \textquote{classical} warp factor of target space metric is given in the large $\phi$ limit by
	\begin{equation}
		g_{cl}(\phi)\approx 1+\exp\left(-\frac{2\phi}{Q}\right)\sum_{i\neq i_0}\frac{|b_{i_0}|^2}{|M_{ii_0}|^2} 
		\label{g_cl_0}
	\end{equation}
	where the radius of the compact dimension
	\begin{equation}
		Q\stackrel{N_0\rightarrow\infty}{\approx}\sqrt{2N_0}
		\label{largeN_Q}
	\end{equation}
	In \eqref{g_cl} we use the rescaled mass differences $M_{ii_0}=-\frac{\beta}{2}(m_i-m_{i0})$ kept finite at $\beta\to 0$, together with
	\begin{equation}
		b_{i_0} \sim \gamma_{i_0}\beta\ \stackreb{\gamma_{i_0} \to \infty,\beta \to 0}{=}\ \text{fixed} \,, \quad
		M \sim m \beta\ \stackreb{m \to \infty,\beta \to 0}{=}\ \text{fixed}.
		\label{limit_hierarchy_1}
	\end{equation}
	It has been shown in \cite{GIMMY} that the equivalence of the Coulomb branch of \wcpN model with \ntwo Liouville theory is actually exact at vanishing bare masses, if the large-$N$ result for the radius \eqref{largeN_Q} is changed for
	\begin{equation}
		Q=\sqrt{2(N_0-1)}
		\label{QN0}
	\end{equation}
	We follow this reasoning for the mass-deformed theory as well (see \cite{Y_mass_Liouville}), and extend the large-$N_0$ derivation to
	$N_0=2$,  then $Q=\sqrt{2}$ from \eqref{QN0} brings us back to the critical non-Abelian string on the conifold.
	
	Note that the resulting effective action \eqref{S_eff} actually does not depend on particular choice of vacuum  $i_0$ in \eqref{sigma_mod},
	due to	$\mathbb{Z}_K$-symmetry of the proposed special deformation, and metric \eqref{g_cl_0} is simplified to
	\beq
	g_{cl}(\phi)\approx  1+\exp\left(-\frac{2\phi}{Q}\right)\frac{(K^2-1)}{12} \frac{|b|^2}{|M|^2}, 
	\label{g_cl}
	\eeq
	To ensure \eqref{g_cl} we require the complex structure deformation parameters $b_{i_0} \sim \gamma_{i_0}\beta$ on each conifold, associated with a particular $U(2)$ factor in \eqref{M_to_infty}, to be the same
	\beq
	b_i =b, \qquad i=1,...,K=N/2.
	\label{equal_b}
	\eeq
	Then  the sum in \eqref{g_cl_0}  for the special choice \eqref{Zp_mass} and \eqref{equal_b} is easily calculated (see Appendix~\ref{App:Seff}) and gives \eqref{g_cl}.

	 Hence, our interpolation procedure is universal: it relates \(U(N=2K)\) SQCD with a \(U(2)\) theory,
	whose spectrum at strong coupling was found due to relation with \ntwo Liouville, independently on a choice from $K$  factors in \eqref{M_to_infty} we start with. 
	The mass deformation explicitly breaks the 2D conformal invariance, so  the effective action
	\eqref{S_eff} cannot be used directly for string quantization. 
	Instead, the \textquote{classical} warp factor \eqref{g_cl} can be used only as an initial condition for solving the effective supergravity equations \cite{Y_mass_Liouville}.

	\subsection{Solutions of gravity equations and  2D black hole}
	\label{sec:gravity_solution}
	
	The bosonic part of the effective action of the type-II supergravity
	in string frame for the metric and the dilaton is given by
	\beq
	S= \frac1{2\kappa^2}\int d^D x \, \sqrt{-G}\,e^{-2\Phi}\,\left\{ \mathcal{R} + 4G^{MN}\pt_M\Phi\pt_N\Phi 
	+ \cdots \right\},
	\label{gravity_action}
	\eeq
	where $G_{MN}$ is the $D$-dimensional metric ($M,N=1,...,D$) with scalar curvature \( \mathcal{R}\), and  the gravitational coupling 
	can be written as 
$2\kappa^2= \frac{\pi }{2}\, g_0^2\,(\alpha')^{\frac{D}{2}-1}$. 
  As usual in string frame the dependence on string coupling 
\(g_s=e^{\Phi}\) comes from the dilaton exponent, therefore one can absorb the dimensionless \(g_0^2\)
into the dilaton constant, making in \eqref{gravity_action} a substitution:
\beq
2\kappa^2 \to 2\kappa_0^2\equiv \frac{\pi}{2}(\alpha')^{\frac{D}{2}-1}, \quad e^{-2\Phi}\to
e^{-2(\Phi +\delta\Phi_0)}, \quad \delta\Phi_0= \log{g_0},
\label{g_0}
\eeq
where now $2\kappa_0^2$ is solely determined by $\alpha'$ or  string tension~\footnote{We stress again, that in our theory of non-Abelian string both $\alpha'$ and $\kappa_0^2$ are of SQCD scale rather then
string/Plank scale of fundamental string theory, and since our string theory contains no gravity in 4D we cannot  fix numeric constant \(g_0^2\) by comparison with gravitational constant, coming from Newton's law. } \eqref{ten}.

	
	Einstein's equations, which ensure 2D conformal invariance by vanishing of the world-sheet beta-functions,  corresponding to the action \eqref{gravity_action} have the form
	\beq
	\mathcal{R}_{MN}+ 2D_M D_N \Phi =0,
	\label{Einstein}
	\eeq
	and are supplemented by the equation for the dilaton
	\beq
	4G^{MN}\pt_M\Phi\pt_N\Phi -2G^{MN}D_M D_N \Phi +\frac{D-10}{2}=0.
	\label{dilaton_eq}
	\eeq
	Assuming that our target-space is a direct product of the flat 4D Minkowski space and internal 2D space with nontrivial metric,  
	associated with  mass-deformed \ntwo Liouville theory
	\begin{equation}
	ds^2_{{\rm int}} = g(\phi) \left(d\phi^2 + dY^2\right) ,
	\label{int_metric}
	\end{equation}
	so $D=6$ in \eqref{gravity_action}, one can find \cite{Y_mass_Liouville} the metric warp factor and the dilaton, 
	\begin{equation}
		g(\phi)= \frac1{1-e^{- Q (\phi-\phi_0)}},\ \ \ 
		\Phi (\phi)= 
		-\frac{Q}{2} \phi -\frac12\, 
		\log{\left[1-e^{- Q (\phi-\phi_0)}\right]},
		\label{gPhi_sol}
	\end{equation}
	where one has to take critical \(Q=\sqrt{2}\) to match the initial condition \eqref{g_cl}, which is satisfied with the choice of the parameter
	\begin{equation}
		\phi_0 = \frac{1}{Q}\log\frac{(K^2-1)|b|^2}{12|M|^2}, 
		\label{phi0KM}
	\end{equation}
	which becomes dependent  on the rank of gauge group $N=2K$.
	
	At $M\to \infty$ solution \eqref{gPhi_sol} gives
	flat metric with linear dilaton \eqref{linear_dilaton}~\footnote{Equation \eqref{dilaton_eq} is however satisfied only if $Q=\sqrt{2}$ is matched with $D=6$.}.
	Note  also that the first nontrivial term in the expansion of the warp factor \eqref{gPhi_sol} at large $\phi$ gives 
	deformation
	\beq
	(\pt_{z}\phi - i\pt_{z}Y) (\pt_{\bar{z}}\phi + i\pt_{\bar{z}}Y)  \,e^{- Q \phi}.
	\label{non-chiral_deform}
	\eeq
	of free action, by marginal operator with $j=-1$, $m=0$, which is  the bosonic part of  so-called non-chiral marginal deformation of  \ntwo Liouville theory; see \cite{Nakayama} for a review. 
	
	Hence, the bosonic part of the world sheet action of the mass-deformed Liouville theory is
	\begin{equation}
		S_{\rm ws} =\frac{1}{4\pi}\int d^2 x \sqrt{h}\; \left\{\frac1{2}g(\phi)\left[ (\pt_{\alpha}\phi)^2 + (\pt_{\alpha}Y)^2\right] 
		+ \Phi(\phi)R^{(2)} + L_{int}\right\}
		\label{deformed_Liouville}
	\end{equation}
	where the metric warp factor $g(\phi)$ and the dilaton $\Phi(\phi)$ are given by \eqref{gPhi_sol}, while the Liouville
	superpotential  still takes the form \eqref{Liouville_sup}
	since it  is not modified by the mass deformation \cite{Y_mass_Liouville}. 
	The action \eqref{deformed_Liouville} defines  a continuous family of world-sheet CFT's, parametrized by parameter \eqref{phi0KM}, and they all have the central charge $c_L=3(1+Q^2)= 9$ (for our choice $Q=\sqrt{2}$).
	
	
	It was shown in \cite{IMSY} that the mass deformed string background \eqref{gPhi_sol} is actually equivalent to the ``trumpet'' geometry, $T$-dual to the 2D black hole's semi-infinite cigar (with radial coordinate $0\leq \rho < \infty$ and angular coordinate $\theta\sim\theta+2\pi$)
	\begin{equation}
	S_{\rm BH} = \frac{k}{4\pi}\int d^2 x \sqrt{h} \left\{(\pt_{\alpha} \rho)^2 +\tanh^2{\rho} \, (\pt_{\alpha}\theta)^2 \right\}
	+ \frac{1}{4\pi}\int d^2 x\sqrt{h}\, \Phi(\rho) R^{(2)}
	\label{BH}
	\end{equation}
	with the dilaton 
	\begin{equation}
	\Phi(\rho)= \Phi_0 -\log{\cosh{\rho}}
	\label{dilaton_cigar}
	\end{equation} 
	The supersymmetric version of the action \eqref{BH} classically corresponds to the 
	SL($2, \mathbb{R}$)/U(1) coset theory \cite{GVafa,GivKut,MukVafa,OoguriVafa95} 
	with the level  \eqref{k_Q} of supersymmetric Ka\v{c}-Moody algebra.

Note that  the target space geometry \eqref{BH} in the supersymmetric case (in contrast to the bosonic 2D black hole) does not receive the 
stringy $\alpha' \sim 1/k$ corrections (see review \cite{Mertens}) therefore, the effective action \eqref{gravity_action} we used in this section is a valid approximation. 
	

	\subsection{Mass deformed Liouville theory}
	\label{sec:superpot+mass}

	As is commonly believed, \ntwo Liouville theory has mirror description \cite{HoriKapustin} in terms of the \ntwo  supersymmetric version of  2D  black hole \eqref{BH} with the Ka\v{c}-Moody level $k =\frac{2}{Q^2}$, which determines 
	the radius of the cigar $R=\sqrt{2k}$ at $\rho\to\infty$, see \eqref{BH}. 
	The constant $\Phi_0$ in the expression \eqref{dilaton_cigar} (the value of the dilaton at the horizon $\rho=0$) is given by
	\beq
	\Phi_0 =  \Phi_0^{(b)} +\log{g_0}, \qquad  \Phi_0^{(b)}= -\frac1k\,\log{|b|}
	\label{Phi_0_b}
	\eeq
	in terms of the coefficient  $b$ (modulus of the conifold complex structure) in front of the superpotential by ``standard Liouville'' argument with the shift 
	\begin{equation}
			\phi\to\phi-Q\log|b|\equiv \phi -\phi_{\text{wall}}
			\label{L_wall}
	\end{equation}
	where $\phi_{\text{wall}}$ determines the position of the ``Liouville  wall''.
In \eqref{Phi_0_b} we  also take into account the shift \eqref{g_0} of the dilaton related to the gravitational coupling.
	
	As was shown yet in \cite{Wbh},  the constant $\Phi_0$ \eqref{dilaton_cigar} determines the ADM mass  of the black hole
	\beq
	M_{BH}  = \frac{Q}{2}\, e^{-2\Phi_0} =\frac{1}{ \sqrt{2k}}\, e^{-2\Phi_0}.
	\label{M_BH}
	\eeq
	One can consider the compact dimension around the cigar in \eqref{BH} as a thermal circle,
	where \(T=\frac1{2\pi R}=\frac{Q}{4\pi}\) is  the Hawking temperature.
	With this interpretation \eqref{M_BH} can be also understood as a definition of the black hole entropy~\footnote{Note, that this 2D temperature appears only due to the existence of compact direction in the Euclidean internal part \eqref{int_metric} of our string background, and has nothing in common with real temperature of 4D theory. We will also see in Sec.~\ref{sec:omega} that \eqref{dil_ent} is only a part of the whole black hole entropy, surviving at \(k\to\infty\).}
	\begin{equation}
		\label{dil_ent}
		s_{BH}  = \frac{M_{BH} }{T} = 2\pi e^{-2\Phi_0} 
	\end{equation}
	%
	For the supersymmetric black hole, associated with \ntwo Liouville theory we have therefore
	\beq
	M_{BH}^{(b)}  =  \frac{Q}{4\pi} s_{BH}^{(b)} = \frac{Q}{2}\, \frac{1}{g_0^2}\,|b|^{2/k}.
	\label{M_BH_b}
	\eeq
	due to \eqref{Phi_0_b}.
	
	On the other hand, as was discussed in the previous section, the mass-deformed world-sheet background \eqref{gPhi_sol} is \emph{T}-dual to \emph{the same} theory \eqref{BH} with the same $k=2/Q^2$, where now, again taking into account \eqref{g_0}, the dilaton constant 
	\beq
	\Phi_0 =  \Phi_0^{(M)} +\log{g_0}, \qquad \Phi_0^{(M)}= -\frac{Q}{2}\,\phi_0=-\frac{1}{2}\log\frac{(K^2-1)|b|^2}{12|M|^2} 
	\label{Phi_0_M}
	\eeq
	is determined by location of the singularity \eqref{phi0KM} in the solution \eqref{gPhi_sol}, i.e. is actually determined by the rank of gauge group and mass deformation scale $M$.
	The associated black hole mass and entropy is therefore~\footnote{Expressions \eqref{Phi_0_M} and \eqref{M_BH_M} are just generalizations to arbitrary $K$ of the $K=2$ results, obtained in \cite{IMSY}.}
	\beq
	M_{BH}^{(M)}  =  \frac{Q}{4\pi} s_{BH}^{(M)} = \frac{Q}{2}\frac{ (K^2-1)|b|^2}{12\,g_0^2|M|^2}.
	\label{M_BH_M}
	\eeq
	It was conjectured in  \cite{IMSY} that when
	both marginal deformations are switched on (see \eqref{deformed_Liouville}),  one still gets the same dual black hole theory \eqref{BH} with only modified
	``total''  value \(\Phi_0\) of the dilaton constant. We proposed that
	the total mass/entropy of the black hole upon switching on both deformations is just the sum~\footnote{As was already pointed out in \cite{IMSY}
		one can only conjecture this formula in nonlinear 2D theory \eqref{deformed_Liouville}, and any other function with each
		term in the r.h.s. of \eqref{phi_total} dominating at large negative values of \eqref{Phi_0_b} or \eqref{Phi_0_M}, say tropical sum
		\(	e^{-2\Phi_0} = e^{-2\Phi_0^{(b)}}\oplus e^{-2\Phi_0^{(M)}} = \max(e^{-2\Phi_0^{(b)}},e^{-2\Phi_0^{(M)}})\), is equally possible.} 
	of the corresponding masses/entropies \eqref{M_BH_b} and \eqref{M_BH_M}, i.e.
	\begin{equation}
		\exp\left(-2\Phi_0^{\text{total}}\right)= \frac{1}{g_0^2}\left\{\exp\left(-2\Phi_0^{(b)}\right)+ \exp\left(-2\Phi_0^{(M)}\right)\right\},
		\label{phi_total}
	\end{equation}
	which gives for  the total mass/entropy  of the black hole in the theory with both deformations,
	\begin{multline}
	\frac{M_{BH}^{\text{total}}}{Q/2} =  \frac{1}{2\pi} s_{BH}^{\text{total}} =
	\frac{1}{g_0^2}\left(|b|^{2/k} +  \frac{(K^2-1)|b|^2}{12|M|^2}\right)\approx
\\
 \stackrel{K\gg 1}{\approx}\frac{1}{g_0^2}\left( |b|^{2/k} +  \frac{N^2|b|^2}{48|M|^2}\right)
	\label{M_BH_total}
	\end{multline}
	Let us stress here again, that in the world sheet theory on the  cigar \eqref{BH} parameter $k$ (fixed to be $k=1$ for $Q=\sqrt{2}$ for critical string), which determines the spectrum of string states via 2D conformal dimensions \eqref{mass_shell}, does not depend on the rank $N=2K$ of the gauge group in  $U(N)$ SQCD. Dependence on the gauge group comes from the dilaton constant $\Phi_0$, related to the black hole mass/entropy \eqref{M_BH_total}, or to the string coupling. 
	It means that the hadron spectrum  of 4D \ntwo SQCD with $U(N)$ gauge group which emerges as we reduce the  mass parameter $M$  does not depend on $N$ and is  still given by \eqref{tachgrmass}, but the multiplicity of hadron states is expected to grow with $N$. Below in Sec. \ref{SQCD_side} we
	use the field theory arguments  to show that this surprising behavior is quite reasonable in QCD with \ntwo supersymmetry.

	\subsection{\ntwo SQCD: stringy versus Higgs phase}
	\label{sec:string_vs_Higgs}

	As we discussed in the Introduction the phase diagram of \ntwo SQCD with even $N$ and $N_f=2N$ has two phases: the Higgs phase at weak coupling and the stringy phase on the non-perturbative Higgs branch, where the massless stringy baryon $b$ develops VEV. This branch is developed at SQCD ''super strong'' coupling point $\tau_{SW}=1$, which corresponds to $\beta=0$ in the world sheet model \cite{ISY_b_baryon}, see Fig.\ref{fig:phase_diagram}. This is also 
	a point of transition from the resolved conifold with $\beta\neq 0$, $\langle b \rangle =0$ to the  deformed conifold with $\beta = 0$, 
	$\langle b \rangle \neq 0$.
	
	Now let us show that these two phases of 4D SQCD are not analytically connected. The point is that although both phases do have massless states, these states have different quantum numbers with respect to the global symmetry group  \eqref{c+f}.
	
	The massless states in the Higgs phase are ``extra'' quark flavors  $q^{kA}$, $\tilde{q}_{Ak}$  with $A=(N+1),...2N=N_f$, see Sec. \ref{sec:SQCD} and \cite{SYrev} for more details.  Their masses
		are given by  $|m_k -m_A|=|m_k -\tilde{m}_{A-N}|$.
		In the limit of interest, when all $m_k,m_A$ vanish, the hypermultiplets $q^{kA},\tilde{q}_{Ak}$ become exactly massless.
		Their masses are BPS-protected, and they remain exactly massless even if we increase the coupling $g^2$,  going toward the strongly coupled region. In particular, these are the BPS states that belong to short \ntwo multiplets (hypermultiplets), and their masses  are not affected by the Higgs mechanism, which gives non-BPS masses of order $\sim g\sqrt{\xi}$ to gauge fields and the first $N$ quark flavors upon their condensation, see  \eqref{qvev} and  \cite{SYrev}  for a review.
These quarks are screened in the Higgs phase and  belong to the bifundamental representations
\beq
(\bf{\bar{N}, N}), \qquad  (\bf{N, \bar{N}})
\label{bifundamental}
\eeq
of the two $SU(N)$ groups in \eqref{c+f} since their color index $k$ gets ''locked'' to a flavor index from the set $B=1,\ldots,N$.

From the point of view of the world sheet theory (see Sec. 2.2.1), the 4D bifundamental quarks are imprinted (due to the 2D-4D correspondence) as the size moduli $\rho^{j}$, which have the same masses $|m_k-\tilde{m}_{A-N}|$ for $j=A-N$ and also become  exactly massless in the limit $m_A \to 0$. 
		In terms of the 2D coupling constant, this phase corresponds to $\beta \neq 0$.

Instead, in the stringy phase, the bifundamental quarks discussed above become heavy and decouple.  This phase corresponds to the point $\beta = 0$ in the 2D theory on the string; at this point, a Coulomb branch in the world sheet theory  is developed, where the 2D field $\sigma$ acquires an arbitrarily large VEV, making the fields $\rho^j$ massive.  This  signals that the corresponding 4D states (namely, ``extra''  quarks with flavor index $A=(N+1),\ldots,2N$) also become heavy.

The only massless state in the stringy phase is the  lightest baryon, which develops a VEV  $\langle b \rangle \neq 0$ on the non-perturbative Higgs branch. This state corresponds to the complex structure modulus $b$ of the conifold (see \eqref{deformedconi}) and therefore, as any other Calabi-Yau modulus, is associated  with a massless state in 4D, see  \cite{GreenSchWitt}. 
		This argument guarantees that the mass of the $b$-baryon  is exactly zero throughout the entire stringy phase (the order parameter parametrizing this phase is the VEV of $b$). In terms of the  world sheet Liouville  theory, this baryon is a string state with 
		$j=-1/2$, $m=\pm 1/2$, see Sec. \ref{sec:hadrons_from_primary}, and 
		it belongs to the \(\frac12 N(N-1)\)-dimensional antisymmetric representation
\begin{equation}
	\Yvcentermath1
	\yng(1,1)
	\label{b_reps}
\end{equation}
of $SU(N)$  (equally charged  w.r.t. both 
		$SU(N)$ factors of the global group  \eqref{c+f}), where we denote $SU(N)$ representations by their Young diagrams,  see 
		Sec.~\ref{sec:su_2K_spectrum} below. Clearly, \eqref{bifundamental} and \eqref{b_reps} are different, so the two phases cannot be analytically connected. They are separated by a phase transition, which in the world-sheet theory on the non-Abelian string is seen as a conifold transition.

As already mentioned in the Introduction, we also have a dramatic qualitative distinction between the massive spectrum in the two phases.  In the Higgs phase at weak coupling, the physical spectrum is formed by a finite
		number of screened quarks and Higgsed gluons together with their superpartners. In contrast, the stringy phase has infinite towers of
		4D hadron  states \eqref{tachgrmass} typical of the spectrum of string theory. The higher energy states of this spectrum cannot be interpreted as composites  of a finite number of the lightest states.
		This  qualitatively resembles 
		the hadron spectrum in real-world  QCD.

	\section{Density of states and black hole entropy}
	\label{sec:omega}

	
	As was discussed above, the string spectrum does not change as we reduce the mass deformation parameter $M$, but we expect the number of states (i.e. the multiplicity) on each energy level to increase. New states, which were separated by infinite barriers in the limit $M\to\infty$, emerge as we reduce $M$, since the system goes through a certain number of crossovers, similar to walls of
	marginal stability for BPS states, where multiplicity of states jumps. 
	
	In this section we are going to confirm these expectations, first calculating the black hole entropy from the string theory side, and then compare the result to the field theory calculation done in the large $N$ limit. It extends the analysis of \cite{IMSY} from $N=4$ colors to arbitrary even $N$, which  allows us to gain some additional insight.
	
	\subsection{Black hole entropy and spectral density}
	\label{ss:BH_entropy}
	
	To calculate the number of states 
	one can interpret a compact 
	dimension around the cigar as thermal cycle with temperature $T=(2\pi R)^{-1}$, given by the asymptotic cigar radius $R=\sqrt{2k}$, and calculate the entropy of 2D black hole. We stress again, that our real time is one of the 4D Minkowski coordinates and has nothing to do with the compact dimension around the cigar, but we use this  trick with the Euclidean 2D black hole to estimate the  number of states in 4D theory.

	However, it turns out that our value $k=1$ for 2D black hole theory exactly corresponds  
	to the Hagedorn temperature \cite{Hagedorn}.
	At Hagedorn  temperature $T_H$ and  beyond the high-energy levels are no longer suppressed by the Boltzmann factor $\exp{(-E/T)}$
	due to the exponential growth   of the density of string states
	\beq
	\omega(E)\ \stackreb{E\to\infty}{\sim}\  E^\alpha e^{\frac{ E}{T_H}}
	\label{density_states}
	\eeq
	(we come back to the power factor $E^\alpha$ and discuss it in detail below), and the partition function 
	\beq
	Z = \int_{0}^{\infty} dE \,\omega(E)\, e^{-\frac{E}{T}}
	\label{statsum}
	\eeq
	becomes divergent~\footnote{Similarly \cite{FM}, the string theory Green functions can acquire singularities in a finite space-time domain (of the order of the string length).}  at $T \geq T_H$. 
	
	For black holes the Hagedorn behavior is related to the black hole/excited strings transition  \cite{Susskind,HorowPolch}, and in particular the expression for entropy \eqref{dil_ent} works only at low temperatures, where one has a well-defined black hole geometry with small $\alpha'$ corrections ($\alpha' \sim 1/k$ for our theory, and they are actually absent in supersymmetric case). As we  increase the temperature, the theory enters a strong coupling regime, and at certain critical value the black hole turns into excited string. This type of  behavior was found also for our 2D black hole  with  asymptotically linear dilaton \eqref{BH}  in \cite{GivKutRabin}. Moreover, it is argued in \cite{GivKutRabin} that the Hagedorn temperature
	is given by
	\begin{equation}
		T_H= \left.T\right|_{k=1} = \left.\frac{1}{2\pi\sqrt{2k}}\right|_{k=1} = \frac{1}{2\pi\sqrt{2}}
		\label{TD_beta}
	\end{equation} 
	i.e. exactly corresponds to the value $k=1$ we are interested in for the critical non-Abelian string in 4D SQCD.
	
	In Euclidean formulation, the Hagedorn behavior is described by thermal scalar -- a winding string mode around the thermal circle \cite{AtickWitt}. 
	As the temperature approaches $T_H$, this mode becomes massless (and presumably tachyonic at $T>T_H$) leading to an instability, see also
	\cite{GivKutRabin,Kut_wind_condens} and \cite{KazakKostovKut} for the nonsupersymmetric version.

	The thermal scalar for the 2D black hole \eqref{BH} is just our $b$-baryon,
	associated with the conifold complex structure $b$, being a massless state in 4D SQCD.  
	It corresponds to the minimal winding state \cite{IMSY},
	i.e. its quantum numbers are
	\beq
	m=\pm \frac{k}{2}, \qquad j=-\frac{k}{2}. 
	\label{j_m_b}
	\eeq
	Its contribution to the entropy
	of the 2D black hole (normalized to the unit 4D volume) was computed~\footnote{It comes from the effective action 
		\beq
	S_{\rm baryon}= \frac1{2\kappa^2}\int d^D x \, \sqrt{-G}\,e^{-2\Phi}\,\left\{-G^{MN}\pt_M \bar{\mathcal{T}}_b\pt_N \mathcal{T}_b 
	+ |\mathcal{T}_b|^2\right\}
	\label{tachyon_action}
	\eeq
for the thermal scalar in our background, to be added to \eqref{gravity_action}.} 
	in \cite{IMSY}:
	\begin{equation}
		s = \frac{e^{-2\Phi_0^{\text{total}}}}{k-1} = 
		\frac{\sqrt{2k}}{k-1}M_{BH}^{\rm total}
		\label{b_entropy}
	\end{equation}
	following the calculation from \cite{Malda_BH_hagedorn} in nonsupersymmetric case.
	
	In contrast to \eqref{dil_ent}, becoming now (unlike the region \(k\to\infty\)) the subleading contribution into entropy, expression \eqref{b_entropy} takes into account the contribution of thermal scalar (leading at $T\to T_H$, see \cite{AtickWitt}) and diverges~\footnote{This divergence comes from the infrared region $\rho\to\infty$, due to the non-normalizable behavior of the thermal scalar near the Hagedorn point  $k\to 1$.} at critical \(k\to 1\).
 However, it is also proportional to the total ADM mass 
	\eqref{M_BH_total}: at large $M \to \infty$ it does not feel the mass deformation parameter, $s\ \stackreb{M\to\infty}{\sim}\ |b|^2(k-1)^{-1}$,
	but small $M $ enhances the divergence in \eqref{b_entropy}
	\beq
	s\ {\approx}\ \frac{1}{k-1}\,\frac{(K^2-1)|b|^2}{12\,g_0^2|M|^2}
	\label{b_entropy_mu}
	\eeq
	and, moreover, makes it dependent on the rank of gauge group.
	

	
	Formulas \eqref{b_entropy}, \eqref{b_entropy_mu} give actually not only the total entropy or the total number of states, they also allow to understand that  
	multiplicity at each energy level (or  the number of states with fixed baryonic charge $B$) increases dramatically as we reduce $M$. To estimate
	the number of states at each  high-energy level (filled now since we meet the Hagedorn behavior) or hadron mass, given by \eqref{tachgrmass} (and depending on the baryon number $B$ via \eqref{m-baryon}), let us first compute the integral \eqref{statsum} for the spectral density \eqref{density_states}. We arrive at
	\begin{equation}
		Z=[2\pi(R-R_H)]^{-(\alpha+1)}\Gamma(1+\alpha),
		\label{alpha_Z}
	\end{equation}
	so calculating the entropy   
	\begin{equation}
		s=\left( 1- R\frac{\pt}{\pt R}\right) \log{Z}\ \stackrel{k\rightarrow 1}{\approx}\ \frac{2(\alpha+1)}{k-1}
		\label{alpha_s}
	\end{equation}
	and  comparing  it with \eqref{b_entropy}, \eqref{b_entropy_mu} allows us to fix the parameter $\alpha$ in \eqref{density_states}. It determines the  growth of the multiplicity of states in the subleading order at large $E$. 
	Namely, comparing \eqref{b_entropy_mu} and \eqref{alpha_s} gives
	\begin{equation}
		\alpha+1=\frac1{24 g_0^2} \frac{(K^2-1)|b|^2}{|M|^2}
	\end{equation}
	Plugging this back into \eqref{density_states}, we finally arrive at the spectral density
\begin{multline}
 \omega(E)\approx \exp\left\{\frac{E}{T_H} + \frac{(K^2-1)|b|^2}{24\, g_0^2|M|^2}\log E +\cdots \right\}\approx
\\
 \stackreb{K\gg 1}{\approx}  \exp\left(\frac{E}{T_H} + \frac{N^2}{96} \frac{|b|^2}{\Lambda^2}\log E +\cdots \right)
	\label{omega_subl}
\end{multline}
where we have introduced $\Lambda \equiv g_0|M|$.
	%
	Here and below we use the large $N$ limit, explicitly for high energies we assume
	\begin{equation}
		E \gg N \gg 1
		\label{E_gg_N}
	\end{equation}
	in the units where $\alpha'=2$, see \eqref{a2} in Sec.~\ref{sec:hadrons_from_primary}.
	
To avoid the  
singularity at $M\to 0$
in the $\log{E}$ term in \eqref{omega_subl}, not expected in 4D SQCD, we require in the limit $M\to 0$
\beq
g_0\to\infty, \qquad M\to 0, \qquad \Lambda =g_0 |M| =\text{fixed}.
\label{lambda}
\eeq
At high energy  the main contribution to the number of states is due to the linear in $E$ term  in the exponent of \eqref{omega_subl},  with the coefficient determined by Hagedorn temperature, which is just a constant \eqref{TD_beta}. However the subleading logarithmic term  $\sim \log E$ in \eqref{omega_subl} grows as  $N^2$ with the rank of the SQCD gauge group.
The coefficient in front of $\log{E}$ can be also written as
\begin{equation}
	\frac{e^{-2\Phi_0^{\text{total}}}}{2} = \frac1{2 g_s^2}, \qquad g_s=\frac{4\sqrt{3}}{N|b|}\,\Lambda
	\label{npt_dyn}
\end{equation}
showing, that this is a non-perturbative contribution in string coupling constant. We also see that actually $\Lambda$, introduced in \eqref{lambda}, determines the string coupling for SQCD with \(U(N)\) gauge group~\footnote{Notice, that at \(M\to\infty\) (i.e. for \(U(2)\) gauge theory with massless quarks) the string coupling is rather given by \(g_s\sim g_0/|b|^{1/k}\ \stackreb{k\to 1}{=}\ g_0/|b|\), i.e. large \(g_s\) is suppressed here only by ``Liouville wall'' \eqref{L_wall}.}. If $\Lambda \sim 1$ (in our units \(\alpha'=2\), see \eqref{ten})
string coupling $g_s$ is small at large $b$ (this is enhanced in the large $N$ limit) so the classical approximation
to string theory we used in Sec.~\ref{sec:gravity_solution} is indeed reliable.

Below we are going to discuss in detail the field theory side and find a quantitative agreement of the $N$ dependence of the $\log E$ term in \eqref{omega_subl} with the result from the SQCD
side in the large $N$ limit.

\subsection{Field theory computation}
\label{SQCD_side}

The string argument above says that the spectrum of the 4D \ntwo SQCD (by spectrum we mean solely the 4D energy levels/masses of states, see \eqref{tachgrmass}) does \emph{not} change as we reduce $M$ interpolating from U(2) to $U(2K)$ theories (with $N_f=2N=4K$ flavors). The \emph{only} effect is growth of the multiplicity of states \eqref{omega_subl}.

This result from string theory side seems to run into a contradiction with intuitive expectations
on the field theory side. Naively, as we reduce $m$, one would expect 
emergence of new hadronic states in SQCD with masses of order $m$.
However, it was argued in \cite{IMSY} that the behavior predicted by string theory is actually  natural for \ntwo SQCD.  Below we generalize these arguments to arbitrary even $N=2K$ and estimate the density of hadron states from the SQCD side at large $N$.

Certainly we cannot strictly calculate the hadron masses directly from SQCD, since the theory is in extremely strong-coupling regime $g^2\to \infty$ (associated with $\beta=0$). 
However, in \ntwo SQCD one can make some plausible assumptions about the dependence of hadron masses on bare quark masses  and quantum numbers of allowed states. 

The hadron mass naturally receives two contributions: one comes from the constituent monopole masses (playing the role of the constituent quarks, see \cite{SYrev}), while the other is from the energy of confining string.
The first contribution is determined by BPS central charges $Z_{BPS}=Z_{BPS}(m_i, \tilde{m}_j)$, given by exact formula 
\cite{SW2,ISY_b_baryon}, while the second one is determined solely by string tension \eqref{ten}.
We have proposed therefore in \cite{IMSY} the following formula
\beq
m_H^2 = m^2_{\text {non-BPS}}(\xi) + |Z_{BPS}(m_i, \tilde{m}_j)|^2.
\label{BPS+non-BPS}
\eeq
for hadron masses, where the BPS part 
\begin{equation}
	\label{Zmq}
	Z_{BPS} = i \vec{m}_n \vec{q}_n - i \vec{m}_\rho \vec{q}_\rho 
\end{equation}
can be found using the 2D-4D correspondence, briefly reviewed in Sec.~\ref{sec:wcp}.
The point is that the BPS spectrum of 4D SQCD in the quark vacuum \eqref{qvev} at $\xi=0$ actually coincides with the BPS spectrum of the 2D world-sheet \wcpN theory of non-Abelian vortex. Hence in \eqref{Zmq} \(\vec{m}_n = \{m_1,\ldots,m_N\}\) and  \(\vec{m}_\rho = \{\tilde{m}_1,\ldots,\tilde{m}_N\}\) are the masses of \(n\) and \(\rho\) fields (they coincide with the masses of $2N$ flavors of quarks, see Sec.~\ref{sec:wcp}). 
Moreover, we introduce the \(\mathfrak{gl}_N\supset \mathfrak{u}_N\) charges (they form the basis of the lattice \(\mathbb{Z}^N\subset \mathbb{R}^N\))  \(\vec{q}_n\) and
\(\vec{q}_\rho\)  (instead of \(\mathfrak{sl}_N\supset \mathfrak{su}_N\) charges), since
both $n^i$ and $\rho^j$ fields transform under the fundamental representation \(\mathbf{N}\) of \(\mathfrak{gl}_N\).

For example, the quark $q^{iA}$ with the flavor $A=N+j$ has the same mass $|m_i-\tilde{m}_j|$ as  the gauge-invariant ``meson'' $w^{ij}=n^i\rho^j$ 
with the charges
\begin{equation}
	\vec{q}_{n^i}=\vec{e_i},\quad \vec{q}_{\rho^j}=\vec{e_j},\ \ \   \quad (\vec{e}_i)_j\equiv\delta_{ij}, \quad i,j=1,\ldots,N
	\label{fund_gln}
\end{equation}
following from  
\begin{equation}
	Z_{n^i\rho^j}=i(m_i-\tilde{m}_j),
	\label{Z_pert}
\end{equation}
after substituting \eqref{fund_gln} into the BPS formula \eqref{Zmq}.

Actually the mass formula \eqref{BPS+non-BPS} combines  evidence both from field theory and string theory sides. In particular, \ntwo supersymmetry ensures that lower bounds $|Z_{BPS}(m_i, \tilde{m}_j)|$ for masses of 
all physical states are given by corresponding  central charges of supersymmetry algebra,
so the only assumption in \eqref{BPS+non-BPS} is that the non-BPS part depends only on string scale $\alpha'$ (or $\xi$) and the baryonic charge
\begin{equation}
	B=\pm \sum_j q_j
	\label{Bcharge}
\end{equation}
where \(\vec{q} = \vec{q}_n = \vec{q}_\rho\).
At weak coupling this assumption is supported by observation, that for all perturbative states $m^2_{\text {non-BPS}}(\xi) \sim g^2\xi$ does not depend on quark masses at the leading order in $g^2$,  see \cite{SYrev} for details. At strong coupling the same assumption is supported by our string theory calculation, which shows that hadron spectrum does not depend on the mass deformation parameter $m$.

\subsubsection{$SU(2)$ case}

In order to determine the BPS masses of the non-perturbative states, we need to write down their global charges and then use Eqs.~\eqref{BPS+non-BPS} and \eqref{Zmq}.
Let us start from the limit $M\to\infty$, where the non-Abelian vortex supports the \wcpt model with 
the global symmetry group \eqref{c+f} given by~\footnote{Each $SU(2)$ factor here is broken down to U(1) by the non-zero quark mass differences.}
\beq
SU(2)\times SU(2)\times U(1)_B.
\label{glob_sym_N=2}
\eeq
The lightest of the non-perturbative states is massless baryon, associated with the conifold complex structure parameter $b$ on the CY side. 
Eq. \eqref{deformedconi} shows that 
the $b$-baryon  transforms with respect to \eqref{glob_sym_N=2} as \(\det w^{ij}\), namely, it is a singlet with respect to both $SU(2)$ factors and has baryonic charge \eqref{Bcharge} $B=2$.  
The $\mathfrak{u}_2$ charges for this state are found from \eqref{fund_gln}, i.e.
\begin{equation}
	\vec{q}_{b}=(1,0)+(0,1)=(1,1),
\end{equation} 
so that both \(SU(2)\) charges \((\vec{q}_{b})_1 -(\vec{q}_{b})_2=0\) vanish, while \(B=(\vec{q}_{b})_1+ (\vec{q}_{b})_2=2\)
and \eqref{Zmq} gives the corresponding central charge  \cite{ISY_b_baryon}
\begin{equation}
	Z_{b}=i(m_1+m_2-\tilde{m}_1-\tilde{m}_2).
	\label{Z_b}
\end{equation}
Note that within our interpolation procedure the mass of this state in the \wcpt model vanishes even at finite $M$ because $\tilde{m}_i=m_i$.

All massive string states in $N=2$ theory are also singlets with respect to both $SU(2)$ factors in \eqref{glob_sym_N=2} (see Sec.~\ref{sec:hadrons_from_primary}), so that they differ only by their baryonic charges $B$. From the point of view of transformation properties these states should behave as  powers of $\det w$ in order to 
exist on the deformed conifold~\footnote{We call this below as ``\textit{conifold rule}''.}, see \eqref{deformedconi}.
Therefore, their central charges are just multiples of  \eqref{Z_b}, and the BPS part of the mass for these states always vanish. They become massive only due to the first non-BPS term in 
\eqref{BPS+non-BPS}, depending only on the string tension $\tau=2\pi\xi$.

\subsubsection{$SU(2K)$ with $K\geq 2$  }
\label{sec:su_2K_spectrum}

In general $SU(N)$ case (with even $N$) some extra points have to be clarified.
The global symmetry group \eqref{c+f}
\beq
SU(N)_{C+F}\times SU(N)\times U(1)_B=SU(N)_{n}\times SU(N)_{\rho}\times U(1)_B
\label{glob_sym_N}
\eeq
at $M\to\infty$ is broken down to
\begin{equation}
	SU(N)\rightarrow SU(2)_{1,2}\times SU(2)_{3,4}\times\ldots \times SU(2)_{N-1,N},
	\label{SU(N)_breaking}
\end{equation}
at each of the $SU(N)$ factors in \eqref{glob_sym_N}.
Following \cite{IMSY}, in order to determine the allowed quantum numbers of the nonperturbative stringy states w.r.t. to the group \eqref{glob_sym_N},
we propose the following generalization of the ``\textit{conifold rule}'':
\begin{enumerate}
	\item 
	First, we assume that 
	all allowed  stringy states in the limit  $M\to\infty$ are singlets with respect to all
	$SU(2)_{2i-1,2i},\,i=1,\ldots,K$ factors in \eqref{SU(N)_breaking} for both $SU(N)$ groups associated with $n$ and $\rho$ fields.   
	\item 
	Second, a generalized ``conifold rule'' requires that the charges of stringy states with respect to the $SU(N)_{\rho}$ and $SU(N)_{n}$ in \eqref{glob_sym_N} coincide, i.e.
	\beq
	\vec{q}_{\rho}=\vec{q}_n.
	\label{conifold_rule}
	\eeq
	\item
	As we reduce $M$ and the global group \eqref{glob_sym_N} is restored, all hadron states 
	should belong to irreducible representations of this group~\footnote{Stronger requirement for all states to  be singlets of the whole $SU(N)$  would kill, for example, the $b$-baryon state, which is expected to survive the transition to a theory with a larger gauge group. In contrast to the \(N=2\) case, the massless $b$-baryon now belongs to a nontrivial (second fundamental)  representation {\tiny\Yvcentermath1$\yng(1,1)=\yng(1)\wedge\yng(1)$} of dimension $\frac12 N(N-1)$ (where {\tiny\Yvcentermath1 $\yng(1)$} denotes the fundamental representation ${\bf N}$). 
		\label{foot_bar}}. To select the allowed representations we 
	take all states, satisfying two requirements above, as highest weight states of corresponding representations.
	
\end{enumerate}
%

The first  rule above determines the form of the highest weight vectors $\vec{q}$ ($\vec{q}\equiv \vec{q}_n=\vec{q}_{\rho}$, see \eqref{conifold_rule}) of allowed irreducible baryonic representations $\lambda=\lambda_{\vec{q}}$. Namely, they should be orthogonal to the roots of corresponding $SU(2)_{2k-1,2k}\subset SU(N)$ subgroups:
\begin{equation}
	\vec{\alpha}_k\vec{q}=0,\ \ \ \ \ \vec{\alpha}_k=\vec{e}_{2k-1}-\vec{e}_{2k}, \qquad k=1,\ldots,K.
	\label{baryon_def}
\end{equation}
The solution of \eqref{baryon_def} is obviously given by
\begin{equation}
	\vec{q}=\sum_{k=1}^{K}\tilde{\lambda}_k(\vec{e}_{2k-1}+\vec{e}_{2k}),
	\label{h_weight}
\end{equation}
where $\{\tilde{\lambda}_k\}_{k=1}^K$ is a set of non-negative integers satisfying
\begin{equation}
	\tilde{\lambda}_1\geq\tilde{\lambda}_2\geq\ldots\geq\tilde{\lambda}_K
	\label{lambda_set}
\end{equation}
to ensure that the representation $\lambda$ is irreducible.

The massless baryon (see footnote~\ref{foot_bar}) corresponds to $\tilde{\lambda}_k=\delta_{k1}$, i.e. (at small $M$, when the global group \eqref{glob_sym_N} is restored) it belongs to the antisymmetric representation \eqref{b_reps} which has totally
\beq
\mathrm{dim}\,\lambda_b=\mathrm{dim}\,\lambda_{\vec{e}_1+\vec{e}_2} = \frac{ N(N-1)}{2}
\label{dim_lambda_b}
\eeq
baryonic states with the charges
\begin{equation}
	\vec{q}_b= \vec{e}_i + \vec{e}_j, \qquad i,j=1,...,N, \qquad i<j.
	\label{q_b}
\end{equation}
The non-BPS contribution to \eqref{BPS+non-BPS} vanishes for $j=-1/2$,
$m=\pm1/2$, see \eqref{tachgrmass}  or for the baryonic charge \eqref{m-baryon}, \eqref{Bcharge} \(|B|=4|m|=2\). 
The BPS-part of their masses vanish when all bare masses $\mathbf{m}=0$, and actually on the whole trajectory of our mass deformation 
\eqref{mass_choice} due to \(m=\tilde{m}\) and  \(\vec{m}_n = \vec{m}_\rho \).
For arbitrary quark masses this multiplet splits with the central charges given by the exact formula \eqref{Zmq},
\beq
Z^{(b)}_{BPS}= i(m_i + m_j -\tilde{m}_i -\tilde{m}_j), \qquad i,j=1,...,N, \qquad i<j
\label{Z_b_N}
\eeq
and baryon masses are given by $m_b=|Z^{(b)}_{BPS}|$, c.f. with \eqref{Z_b} for $N=2$.

All massive  baryonic states (which can be thought of as composites made of $n$ elementary $b$-baryons) belong to the irreducible representations, arising in decomposition of the tensor product 
\begin{equation}
	\Yvcentermath1
	\underbrace{\yng(1,1)\otimes\yng(1,1)\otimes \ldots \otimes\yng(1,1)}_{n}
	\label{long_tens_prod}
\end{equation}
where arbitrary integer $n>1$ is determined by the baryonic charge
\begin{equation}
	|B| = \sum_{i=1}^N q_i = 2\sum_{j=1}^K\tilde{ \lambda}_j = 2n \,.
	\label{B_from_n}
\end{equation}
However, the previous analysis  shows that one should extract from  \eqref{long_tens_prod} the irreducible representations with the highest weights
\eqref{h_weight}, i.e. corresponding to the Young diagrams~\footnote{The total number of cells \eqref{B_from_n} in this diagram is $2n$.}:
\begin{equation}
	\begin{tabular}{cl}
		\multirow{10}{*}{$\yng(12,12,11,11,8,8,6,6,5,5)$} & \multirow{2}{*}{$\Big\}\tilde{\lambda}_1$}\\
		& \\
		& \multirow{2}{*}{$\Big\}\tilde{\lambda}_2$}\\
		&\\
		& \multirow{2}{*}{$\vdots$} \\
		&\\
		&  \\
		& \\
		& \multirow{2}{*}{$\Big\}\tilde{\lambda}_K$}\\
		& \\
		
	\end{tabular}
	\label{dladder}
\end{equation}
having the form of a \textquote{double-step staircase}, with each $i$-th pair of steps having the length of $\tilde{\lambda}_i$.

As was just discussed, within our interpolation procedure due to the conifold rule \eqref{conifold_rule} the BPS contribution \eqref{Zmq} to the masses of all hadrons vanishes, and they become massive due to the first non-BPS term in 
\eqref{BPS+non-BPS}. This ensures that the hadron spectrum does not depend on bare masses $\mathbf{m}$,  which matches the string theory result.
The requirement \(m=\tilde{m}\) is
also important, since it ensures that coupling $\beta$ does not run and no $\Lambda$-scale is generated in \wcpN model so its Coulomb branch can really be reduced to the conformal (deformed) \ntwo Liouville theory.

\subsubsection{High-energy asymptotics}

To compare the multiplicity of allowed hadron states at small $M$ with the string result \eqref{omega_subl} we consider the high energy limit. Notice that at large energies $E$ (where $E$ is the 4D mass of the  state) it follows from \eqref{tachgrmass}-\eqref{m-baryon}  that $E\sim |B|=2n$. Hence,
the conditions \eqref{E_gg_N} turn into
\beq
n\gg N \gg 1,
\label{n_gg_N}
\eeq
in the representation theory context, or in other words, we should consider the \(n\to\infty\) asymptotic at large, but fixed \(N\).

To calculate the dimension of the irreducible representation one can use the Weyl formula:
\begin{equation}
	\mathrm{dim}\,\lambda=\prod_{1\leq i<j\leq N}\frac{\lambda_i-\lambda_j+j-i}{j-i},
	\label{Weyl}
\end{equation}
where $\lambda_i$, $i=1,...,N$ is the length of the $i$-th row.
For the double-step ladder \eqref{dladder} we put $\tilde{\lambda}_k=\lambda_{2k-1}=\lambda_{2k},\,k=1,\ldots,K$ and get
\begin{equation}
	\dim\lambda_B = \prod_{1\leq i<j\leq K}\prod_{\epsilon,\epsilon'=0,1}  \frac{\tilde{\lambda}_i-\tilde{\lambda}_j+2(j-i)+\epsilon-\epsilon'}{2(j-i)+\epsilon-\epsilon'}
	\label{Weyl_d}
\end{equation}
from \eqref{Weyl}.


To estimate the number of baryonic states at high energy, we need to understand  the behavior of the dimension of an irreducible representation in the limit \eqref{n_gg_N}. For generic representations it is easy to calculate  dimensions from scaling properties of the Weyl formula \eqref{Weyl}, the r.h.s. of which scales under $\{\tilde{\lambda}_i \to t\tilde{\lambda}_i \}$ as \( t^{\frac12 N(N-1)}\).
The number of rows in the Young diagram \eqref{dladder} is restricted by large, but fixed \(N\), so at $n\to\infty$ the diagram ``grows to the right'',
and one can take into account this growth by scaling the row lengths as
\begin{equation}
	\tilde{\lambda}_i\to t\tilde{\lambda}_i \,, \quad
	B\to t\sum_i \tilde{\lambda}_i,
	\label{rescaling}
\end{equation}
where $t = O(n) = O(B)$ is an overall scaling parameter, and all $\tilde{\lambda}_i$ are kept finite after rescaling.
%
For generic double-step diagrams \eqref{dladder}, we get ($N = 2K$) after rescaling  of \eqref{Weyl_d}
\begin{equation}
	\dim\lambda_B \simeq \prod_{1\leq i<j\leq K} \frac{t^4(\tilde{\lambda}_i-\tilde{\lambda}_j)}{4(j-i)^2\left(4(j-i)^2-1\right)}\ \stackreb{t\to\infty}{\sim}\  t^{4\frac12 K(K-1)} \sim B^{\frac12 N^2-N} 
	\label{ladder_sca}
\end{equation}
The leading behavior is actually universal \footnote{
	As we already mentioned for generic Young diagrams the scaling gives 
	\begin{equation}
		\dim\lambda \simeq t^{\frac12 N(N-1)} C_\lambda = B^{\frac12 N(N-1)} \tilde{C}_\lambda
	\end{equation}
	where the constant $\tilde{C}_\lambda = O(1)$ at large $n$, and leading asymptotic behavior is the same (see also Appendix \ref{App:scaling_asympt} for some numeric tests of this scaling behavior).
	\label{foot:Ysca}} and is given by
\begin{equation}
	\log\dim\lambda_B \sim \frac{1}{2} N^2 \log n \sim \frac{1}{2} N^2 \log B  \sim \frac{1}{2} N^2 \log E
	\label{ln_dim_leading_from_scaling}
\end{equation}
This is the asymptotics of 
the multiplicity of baryonic states in given irreducible representation with mass $E\sim n\to\infty$ at
large $N$.


The asymptotics \eqref{ln_dim_leading_from_scaling} can be possible corrected by growth of the \emph{number} of allowed irreducible representations
\(p(B,N)\) at \(B\to\infty\). It is well-known however (see e.g. \cite{partitions}), that the number of partitions of \(B\) into up to \(N\) parts $p(B,N)$ can be analyzed from their generating function
\begin{equation}
	\sum_{B\geq 0}p(B,N)x^B = \prod_{k=1}^N\frac{1}{1-x^k}.  
\end{equation}
At large \(B\) with fixed \(N\) the asymptotic~\footnote{As well as asymptotic of partition in exactly \(N\) parts, in contrast with the famous Hardy-Ramanujan asymptotic of the number of all Young diagrams
	\begin{equation}
		p(B)\ \stackreb{B\to\infty}{\sim}\ \frac{1}{4\sqrt{3}B}\ e^{\pi \sqrt{\frac{2B}{3}}}.
	\end{equation}
	which actually leads to the Hagedorn behavior \eqref{density_states} in string theory.} is
\begin{equation}
	\label{pBN}
	p(B,N)\ \stackreb{B\to\infty}{\sim}\ \frac{B^{N-1}}{N!(N-1)!}\sim e^{N\log B}
\end{equation}
just as \(N\)-th power of \(B\). It follows, say, from a standard ``residue argument''
\begin{multline}
	p(B,N) = \res_{x=0}\frac{dx}{x^{B+1}\prod_{k=1}^N(1-x^k)} = 
	\\
	=- \res_{x=1}\frac{dx}{x^{B+1}\prod_{k=1}^N(1-x^k)} + \ldots \sim 
	\\
	\sim - \res_{x=1}\frac{dx}{x^{B+1}(1-x)^N\prod_{k=1}^N(1+x+\ldots +x^k)} + \ldots \sim
	\\
	\left.\sim \frac1{(N-1)!}\frac{d^{N-1}}{dx^{N-1}}\frac{1}{x^{B+1}\prod_{k=1}^N(1+x+\ldots +x^k)}\right|_{x=1} =
	\\
	= \frac{(B+1)\ldots(B+N-1)}{N!(N-1)!}+\ldots\ \stackreb{B\to\infty}{\sim}\frac{B^{N-1}}{N!(N-1)!}
	\label{res_der}
\end{multline}
The number of double-step representations \eqref{dladder} is therefore given by
\begin{equation}
	p(n,K)=p(B/2,K) \sim e^{K\log B} = e^{\frac{N}{2}\log B}
\end{equation}
so that the total number $\#_{\mathrm{lad}}$ of possible double-step ladders \eqref{dladder}
\begin{equation}
	\log \#_{\mathrm{lad}}\approx \frac{N}{2}\log B \sim  \frac{N}{2}\log E.
	\label{ladder_count}
\end{equation}
The coefficient in front of $\log{E}$ is of the order $O(N)$ and does not change the main $O(N^2)$ asymptotics in \eqref{ln_dim_leading_from_scaling}, though corrects the next term, depending on chosen class of Young diagrams.

We  therefore conclude that the group theory  analysis from SQCD side confirms the $N$ dependence of $\log{E}$ term in string theory asymptotics \eqref{omega_subl} at $N\gg 1$. However,  the group theory result \eqref{ln_dim_leading_from_scaling} certainly cannot reproduce the non-perturbative coefficient \eqref{npt_dyn}   in front of $\log{E}$ term, arising from string theory dynamics.

\subsubsection{Example: \textquote{Box} diagrams
\label{ss:box}}

\begin{figure}[h]
	\centering
	\includegraphics[width=0.7\linewidth]{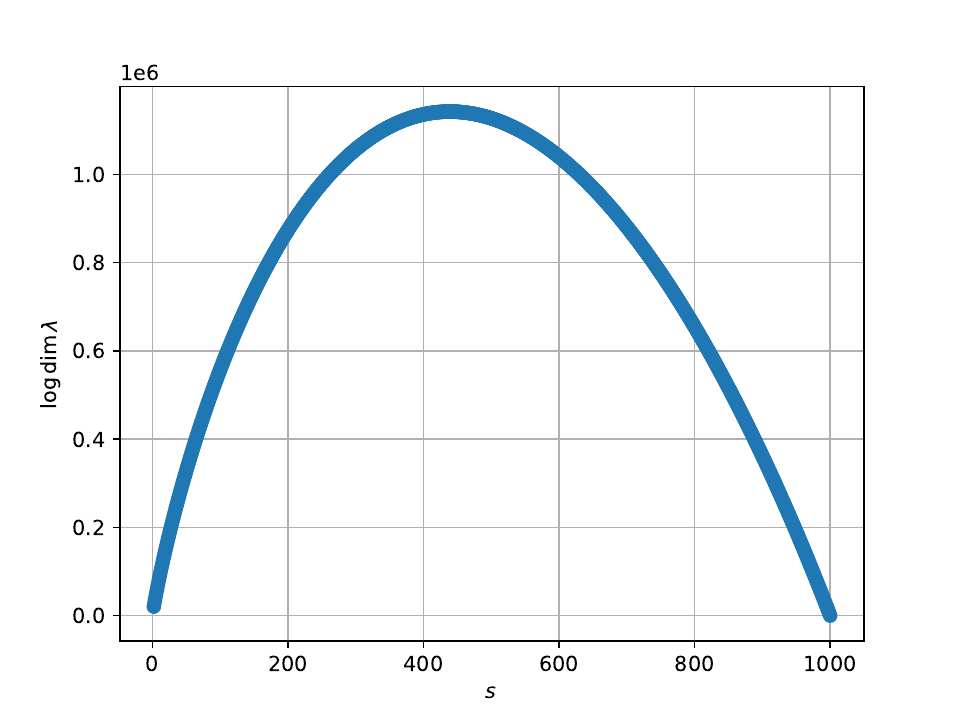}
	\caption{
		Numerical plot of the dependence of $\log \dim\lambda$ on s with $n=10000000,N=1000$ (source: \cite{github}).
	}
	\label{fig:box_lndim_s}
\end{figure}

In order to demonstrate some features of the asymptotic formulas let us present here (see also Appendix~\ref{App:dim_calc}) an explicit example of the 
box-shaped diagrams, defined by 
\begin{equation}
	\begin{cases}
		{\lambda}_1={\lambda}_2=\ldots={\lambda}_s, \quad &\text{for some even $s\in[1,N]$} \\
		{\lambda}_i=0,  \quad &i=s+1,\ldots, N
	\end{cases}
	\label{box_lambda}
\end{equation}
These are not the generic diagrams, and the previous scaling analysis (see \eqref{ladder_sca} and footnote~\ref{foot:Ysca}) is not applicable. Indeed, it is easy to see from the Weyl formula \eqref{Weyl}, that the dimensions of ``box'' representations \eqref{box_lambda} scale under \(\lambda_s\to t\lambda_s\)
at \(t\to \infty\) as \(t^{s(N-s)}\). Hence, when $s$ goes from $1$ to $N$, the dimension $\mathrm{dim}\,\lambda$ first grows and then decreases, see the result of numeric analysis on Fig.~\ref{fig:box_lndim_s}, 
which actually corrects the naive scaling parabola.

For the thin or thick strips with either $s=2$ or $s=N-2$ the  dimension formula \eqref{Weyl} gives
scaling
\begin{equation}
	\log \mathrm{dim}\,\lambda^{(s=2)}\ \stackreb{n\to\infty}{\sim}\  2N\log n,\quad \log \mathrm{dim}\,\lambda^{(s=N-2)} \ \stackreb{n\to\infty}{\sim}\   2N\log n,
\end{equation}
see more details in Appendix~\ref{App:dim_calc}. So, for the diagrams with either \(s\ll N\) or \(s\gg 1\) the leading asymptotics \(N^2\log E\) is absent, and they give contribution of the same order as the overall  number of diagrams \eqref{ladder_count}.

The naive maximum for the ``box'' diagrams is at $s=N/2=K$ (square-shaped diagram), where
\eqref{Weyl} gives
\begin{equation}
	\log \mathrm{dim}\,\lambda^{(s=N/2)} \sim \frac{N^2}{4}\log n. 
	\label{square_diagram_dim}
\end{equation}
We see therefore, that among the box diagrams, which are a special case of \textquote{double-step ladders} for even \(s\), those with \(s\sim N/2\) have the same behavior of asymptotic dimension  
\begin{equation}
	\log \mathrm{dim}\,\lambda^{(s\sim N/2)} \sim N^2\log n
	\label{N_dep}
\end{equation}
as generic diagrams, which match the $\log{E}\approx \log{n}$ term calculated from  the  string theory, see  \eqref{omega_subl}. However, the numerical coefficient in front of $N^2\log n$ is not universal, and cannot be determined just by representation theory analysis.

\section{Massless baryons as Goldstones}
\label{Sec:Goldstones}

As we already discussed in strongly coupled \ntwo SQCD  a new non-perturbative Higgs branch grows up,  where stringy massless baryon $b$ develops VEV (see Introduction and Fig.~\ref{fig:phase_diagram}).  At $M\to\infty$ in the world sheet theory of  non-Abelian string this VEV is related to unique complex structure parameter of  $K=N/2$ conifolds, associated with each of non-interacting $U(2)$ SQCDs, see \eqref{equal_b}. 

As we reduce $M$ the global flavor group \eqref{glob_sym_N} is restored and massless $b$-baryons form
the antisymmetric representation \eqref{b_reps}   of dimension \eqref{dim_lambda_b} (see also footnote~\ref{foot_bar}).
Below we show that these massless states can be viewed as Goldstone
bosons  resulting from the spontaneous symmetry breaking of the flavor  group  by the VEV of $b$.    

The basis vectors \(\{e_i\wedge e_j\}\) (with the weights \eqref{q_b}) in representation {\tiny\Yvcentermath1$\yng(1,1)=\yng(1)\wedge\yng(1)$} 
can be identified with the skew-symmetric matrices \(\{e_i\wedge e_j^T=E_{ij}-E_{ji}\}\). 
In this basis the $b$-baryon VEV at  $M\to\infty$ is given by
\begin{equation}
	\langle b \rangle = b J \,, \quad
	J \equiv \oplus_{j=1}^{K=N/2}\left(
	\begin{array}{cc}
		0 & 1 \\
		-1 & 0
	\end{array}\right)_j \equiv  \left(
	\begin{array}{cccc}
		\epsilon & 0 & \ldots & 0 \\
		0 & \epsilon &  \ldots & 0 \\
		\vdots &  &  \ddots & \vdots \\
		0 &  & \ldots & \epsilon 
	\end{array}
	\right).
	\label{b_VEV}
\end{equation}
It is invariant under \(\mathbb{Z}_K\) and unbroken \(\oplus_{j=1}^{K=N/2} \mathfrak{sl}(2)_j\) action,
since \(\epsilon =\left(
\begin{array}{cc}
	0 & 1 \\
	-1 & 0
\end{array}\right) \) is the \(\mathfrak{sl}(2)\simeq \mathfrak{sp}(2)\)-invariant, and $b$ is a unique (see ~\eqref{equal_b}) complex parameter, which determines the value of the $b$-condensate \footnote{Note that we can consider only one of two $SU(N)$ groups in \eqref{glob_sym_N} due to the  ''conifold rule'' \eqref{conifold_rule}.}.

The baryon VEV \eqref{b_VEV} breaks flavor group \(SU(N)\) down to its (compact) \(Sp(K)\subset SU(N)\) subgroup, which preserves a symplectic\footnote{Note that the symplectic form $J$ from \eqref{b_VEV} can be brought to more conventional form, e.g. $\begin{pmatrix} 0 & I \\ -I & 0 \end{pmatrix}$, by permutation of the basis elements. } 
form $J$, which satisfies \(J+J^T=0\) and \(J^2=-1\).

As mentioned above, $b$ belongs to the second fundamental representation \eqref{b_reps}, implying the transformation law
\begin{equation}
	b'=Ub\,U^T,\qquad U \in \text{$SU(N)$}
	\label{b_trans}
\end{equation}
and VEV \eqref{b_VEV} breaks this symmetry down to Sp($K$).
The \(\mathfrak{sp}(K)\subset \mathfrak{sl}(N)\) subalgebra corresponds to invariant \(\sigma(X)=X\) part of the automorphism 
\begin{equation}
	\sigma(X) = JX^TJ,\ \ \ X\in \mathfrak{sl}(N),
\end{equation}
i.e. to the matrices satisfying
\begin{equation}
	\label{XJ}
	JX+X^TJ=0,\ \ \ X\in \mathfrak{sl}(N)
\end{equation}
whose action preserves \(J\).

The coset space \(SU(N)/Sp(K)\) corresponds to the Goldstone modes, arising due to spontaneous symmetry breaking. 
Apart from that, the VEV of $b$ also breaks the overall $U(1)_B$.
The number of Goldstone modes is, therefore, 
\begin{multline}
\#_{\mathrm{Gold}}=	1 + \dim \mathfrak{sl}(N)- \dim \mathfrak{sp}(K) =
\\
=	1 + (N^2 - 1) -\frac{N(N+1)}{2} = \frac{N(N-1)}{2}
	\label{numGold}
\end{multline}
This confirms our expectations, since \eqref{numGold} exactly coincides with \eqref{dim_lambda_b}. Note that $b$-baryon form a hypermultiplet in \ntwo SQCD \cite{KSYconifold} so the multiplicity  in \eqref{numGold} should be multiplied by four to account for all real scalar superpartners.

Embedding  \(\mathfrak{sp}(K)\subset \mathfrak{sl}(N)\) (and corresponding form of the matrices in coset) can be described explicitly, the simplest way is to use the basis coming from the folding of Dynkin diagrams, see Appendix~\ref{app:goldstone}. Dimension \eqref{numGold} can be also checked numerically, by explicit analysis of the formula \eqref{b_trans}
in its infinitesimal form, see Appendix~\ref{app:num_goldstone}.

\section{Conclusions}
\label{sec:concl}

In this paper we studied the  hadronic phase of strongly coupled 4D $\mathcal{N}=2$ SQCD using
the special quark mass deformation for interpolation from SQCD with $U(2)$ gauge group and $N_f=4$
quark flavors to $U(N)$ SQCD with $N_f=2N$ flavors for arbitrary even $N$. Implying a stringy interpretation of  hadrons of  \ntwo SQCD we found the description of the critical non-Abelian string in terms of mass-deformed  \ntwo Liouville theory. 
Turning on simultaneously the Liouville superpotential and the mass deformation on the world sheet, we used the dual Euclidean 2D black hole picture to find the   mass spectrum of string states. We were also able to find the 
growth of the multiplicity of states at high energies from analysing  the entropy of 2D black hole.

A key outcome is that, while the positions of the stringy energy levels do not change along the interpolation in the mass deformation parameter $M$, the multiplicity on each level grows, and becomes parametrically large as we reduce $M$, as reflected by enhancement of the black hole entropy, cf.\ \eqref{b_entropy_mu}. 

Extracting the distribution of states from the behavior of black hole entropy near the Hagedorn temperature, we fixed the asymptotic spectral density at high energies in the form
\[\omega(E)\sim E^\alpha e^{\frac{ E}{T_H}}\]
with the universal leading Hagedorn exponent and a subleading power correction, whose power depends non-trivially on the deformation data, see \eqref{omega_subl}. The coefficient \(\alpha\sim 1/g_s^2 \sim N^2|b|^2/\Lambda^2\) is non-perturbative in string coupling, determined in this regime by  $N$ and the parameter \eqref{lambda}.
This provides a quantitative prediction for how the multiplicity of baryonic string states depends on the rank $N$ of the  gauge group in SQCD.

We then justified these results directly in the 4D language by analyzing the spectrum of hadrons 
given by Eq.~\eqref{BPS+non-BPS}. It separates the non-BPS contribution given by string theory results \eqref{tachgrmass} from the BPS part, given by exact formula depending on quark masses and global hadron charges with respect to the flavor group \eqref{glob_sym_N}.
The allowed baryons  populate a restricted family of irreducible representations of \eqref{glob_sym_N} characterized by double-step (ladder) Young diagrams. Using the Weyl formula for the dimensions of irreducible representations and calculating the number of allowed Young diagrams, we found the characteristic behavior of the logarithm of the baryon multiplicity to be
\[ \sim N^2\log B \]
at fixed large rank $N\gg 1$ and baryon number $B\to\infty$ \eqref{n_gg_N},
confirming the string theory result  \eqref{omega_subl} for $N^2$-dependence of the $\log{E}$-term.

Finally, we showed that the massless $b$-baryon multiplet in $U(N)$ SQCD can be 
understood as Goldstone modes upon spontaneous breaking of the global flavor group \eqref{glob_sym_N} by $b$-baryon condensate.
The corresponding VEV in the  representation \eqref{b_reps} breaks $SU(N)\times U(1)_B$ down to $Sp(N/2)\subset SU(N)$, yielding exactly $\frac{N(N-1)}{2}$ massless states, in agreement with dimension of the massless baryon representation \eqref{dim_lambda_b}. 

To summarize the dynamics of \ntwo SQCD with $N_f=2N$  flavors we presented its phase diagram, see Fig.~\ref{fig:phase_diagram}. The theory has two phases, the Higgs phase at weak coupling and the hadronic/stringy phase at strong coupling where the massless stringy $b$-baryon develops VEV forming the non-perturbative Higgs branch. These two phases are separated by a phase transition, see Sec.~\ref{sec:string_vs_Higgs}. 

Physics is dramatically different in these two phases.  The hadron phase has  towers of stringy states (see \eqref{tachgrmass}) typical for  the spectrum of the string theory, while in the Higgs phase at weak coupling the physical spectrum is formed by a finite number of screened quarks and Higgsed gluons together with their superpartners.

It would be interesting to sharpen the matching of  multiplicities of hadronic states calculated from string theory and field theory sides beyond the asymptotic regime by incorporating finite-$N$ corrections and to study dynamical observables (e.g.\ correlation functions and decay amplitudes) that probe  the interactions among the resulting stringy hadrons.

\section*{Acknowledgments}

The authors are grateful to  A.~Artemev and A.~Litvinov  for useful discussions of conformal backgrounds, and to M.~Bershtein and E.~Feigin for illuminating some issues in representation theory. 
The work of E.I. was supported in part by U.S. Department of Energy Grant No.~de-sc0011842. 
The work of A.M. was partially supported by the Basic Research Program HSE-BR-2025-84 of HSE University (Sec.~\ref{SQCD_side} and Sec.~\ref{Sec:Goldstones}), and by 
the Russian Science Foundation under the grant 26-11-00342 (Sec.~\ref{Sec:Mass_def} and Sec.~\ref{ss:BH_entropy}). 
The work of G.S. was also partly supported by the Gribov Scholarship for works in the field of theoretical physics.

\appendix


\section{Large-$N$ derivation of effective action}
\label{App:Seff}

We start with the one-loop kinetic term Eq.~\eqref{S_eff_gen}. 
For the special mass choice \eqref{mass_choice} total number $N\equiv N_0 K$ of quark flavors (here we consider large $N_0$, but set $N_0=2$ at the end) form $K$ groups with equal masses in each group.
Then the sum over masses in \eqref{S_eff_gen} can be rewritten as
\begin{equation}
	\sum_{A=1}^{2N}\frac{1}{|\sigma+\frac{m_A}{\sqrt{2}}|^2}=\sum_{A=1}^{2KN_0}\frac{1}{|\sigma+\frac{m_A}{\sqrt{2}}|^2}=2N_0\sum_{i=1}^{K}\frac{1}{|\sigma+\frac{m_i}{\sqrt{2}}|^2},
\end{equation}
and we get
\begin{equation}
	\begin{aligned}
		S_{eff}&= \frac{2N_0}{8\pi}\int d^2 x |\partial_\alpha \sigma|^2  \sum_{i=1}^{K}\frac{1}{\left|\sigma+\frac{m_i}{\sqrt{2}}\right|^2}\\
		&= \frac{2N_0}{8\pi}\int d^2 x \frac{|\partial_\alpha \sigma|^2}{|\sigma+\frac{m_{i_0}}{\sqrt{2}}|^2}\left( 1+ \sum_{i\neq i_0}^{K}\frac{\left|\sigma+\frac{m_{i_0}}{\sqrt{2}}\right|^2}{\left|\sigma+\frac{m_i}{\sqrt{2}}\right|^2}\right)ю 
	\end{aligned}
	\label{seff_details_1}
\end{equation}
Changing of variables according to \eqref{sigma_mod} one gets
\begin{equation}
	\begin{aligned}
		\left|\partial_\alpha\sigma\right|^2 =\frac{1}{Q^2} \gamma_{i_0}^2 e^{-\frac{2\phi}{Q}} ((\partial_\alpha\phi)^2+(\partial_\alpha Y)^2)=
		\\
		=\frac{1}{Q^2} \left|\sigma+\frac{m_{i_0}}{\sqrt{2}}\right|^2 ((\partial_\alpha\phi)^2+(\partial_\alpha Y)^2).
	\end{aligned}
	\label{seff_details_3}
\end{equation}
Furthermore, the sum in Eq.~\eqref{seff_details_1} can be simplified as
\begin{equation}
	\begin{aligned}
		&\sum_{i\neq i_0}^{K}\frac{\left|\sigma+\frac{m_{i_0}}{\sqrt{2}}\right|^2}{\left|\sigma+\frac{m_i}{\sqrt{2}}\right|^2}=\sum_{i\neq i_0}^{K}\frac{\left|\sigma+\frac{m_{i_0}}{\sqrt{2}}\right|^2}{\left|\sigma+\frac{m_{i_0}}{\sqrt{2}}+\frac{m_i-m_{i_0}}{\sqrt{2}}\right|^2}=\sum_{i\neq i_0}^{K}\frac{1}{\left|1+\frac{m_i-m_{i_0}}{\sqrt{2}\gamma_{i_0}}\exp\left(\frac{\phi+iY}{Q}\right)\right|^2}\\
		&\stackrel{\phi\rightarrow\infty}{\approx}\sum_{i\neq i_0}^{K}\frac{\exp\left(-\frac{2\phi}{Q}\right)}{\left|\frac{m_i-m_{i_0}}{\sqrt{2}\gamma_{i_0}}\right|^2}=\exp\left(-\frac{2\phi}{Q}\right)\sum_{i\neq i_0}^{K}\frac{|b_{i_0}|^2}{|M_{ii_0}|^2}, \quad M_{ii_0}=-\frac{(m_i-m_{i_0})\beta}{2}.
	\end{aligned}
	\label{seff_details_2}
\end{equation}
The $\mathbb{Z}_K$ symmetry makes this sum independent of the choice of a particular $i_0$. Indeed,
with explicit set of masses \eqref{mass_choice}-\eqref{Zp_mass} and universal \eqref{equal_b}, we have:
%
%
\begin{equation}
	\sum_{i\neq i_0}^K\frac{1}{|M_{ii_0}|^2}=\sum_{j=1}^{K-1}\frac{1}{|A|^2m^2\left|e^{\frac{2\pi i j}{K}}-1\right|^2},
\end{equation}
where $A = - \beta / 2$.
The r.h.s. can be further rewritten as
%
%
\begin{equation}
	\begin{aligned}
		\frac{1}{2m^2|A|^2}\sum_{j=1}^{K-1}\frac{1}{\left(1-\cos\left[\frac{2\pi ij}{K}\right]\right)}
		=\frac{1}{2m^2|A|^2}\sum_{j=1}^{K-1}\frac{1}{2\sin^2\left[\frac{\pi j}{K}\right]}=
		\\
		=\frac{1}{2m^2|A|^2}\left(\frac{K^2-1}{6}\right)
	\end{aligned}
\end{equation}
%
In the large-$K$ limit, we therefore obtain 
\begin{equation}
	\sum_{i=2}^K\frac{1}{|M_{ii_0}|^2} = \frac{K^2-1}{12|M|^2}\ \stackreb{K\gg 1}{\approx}\ \frac{K^2}{12|M|^2}, \quad |M|^2\equiv \left|\frac{m\beta}{2}\right|^2,
\end{equation}
which gives   \eqref{g_cl}.

\section{Some allowed representations and asymptotics}
\label{App:dim_calc}

\subsection{Box diagrams}

For illustrating the results of Sec.~\ref{sec:su_2K_spectrum}-\ref{ss:box}  we consider the box Young diagram of height $s$
\begin{equation}
	\begin{tabular}{rl}
		
		\multirow{3}{*}{$s=2$}& \multirow{3}{*}{$\xupdownarrow[0.6cm]\yng(15,15)$} \\
		&\\
		&\\
		\multirow{4}{*}{$s=N/2$}& \multirow{4}{*}{$\xupdownarrow[1.5cm]\yng(10,10,10,10)$} \\
		&\\
		& \\
		& \\
		\multirow{6}{*}{$s=N-2$}& \multirow{6}{*}{$\xupdownarrow[2.5cm]\yng(7,7,7,7,7,7)$}\\
		& \\
		& \\
		& \\
		& \\
		& \\
	\end{tabular}
	\label{some_box_diagrams}
\end{equation}
and demonstrate application of the Weyl formula \eqref{Weyl} to computation of their dimensions.
From eq. \eqref{box_lambda} together with $\sum_{i=1}^N\lambda_i=2n$ we have
\begin{equation}
	\lambda_i=\frac{2n}{s},\quad 1\leq i\leq s;\ \ \ \ \ \ \  \lambda_j=0,\quad s< j\leq N
	\label{box_lambda_i}
\end{equation}
Let us start with the $s=2$ case, where only nonvanishing $\lambda_1 = \lambda_2=n$, and the Weyl formula \eqref{Weyl} just gives
\begin{equation}
	\begin{aligned}
		\dim \lambda^{(s=2)}&= \frac{\lambda_1-\lambda_2+2-1}{2-1}\prod_{3\leq j\leq N}\frac{\lambda_1+j-1}{j-1}\times
		\prod_{3\leq j\leq N}\frac{\lambda_2+j-1}{j-1}\\
		&\stackrel{n\gg N}{\approx} 1\cdot(\lambda_1)^{N-2}\lambda_2^{N-2}\stackrel{N\gg 1}{\approx} n^{2N}.
	\end{aligned}
\end{equation}
For the conjugated diagram with $s=N-2$ the dimension of the representation is the same.

As an opposite example, consider the box with exactly $s=N/2$. In this case,
\( \lambda_1=\lambda_2=\ldots\lambda_{N/2}=\frac{4n}{N}\), and plugging this into the Weyl formula yields
\begin{equation}
	\begin{aligned}
		\dim \lambda^{(s=N/2)} =
		\prod_{N/2<j\leq N}\prod_{1\leq i\leq N/2}\frac{\lambda_i+j-i}{j-i}
		\stackrel{n\gg N}{\approx} \prod_{N/2<j\leq N} \left(\frac{4n}{N}\right)^{\frac{N}{2}}\approx n^{\frac{N^2}{4}}.			
	\end{aligned}
	\label{lad_count_sum}
\end{equation}

\subsection{Numeric illustration of scaling formula}
\label{App:scaling_asympt}

\begin{figure}[h!]
	\centering
	\includegraphics[width=0.6\linewidth]{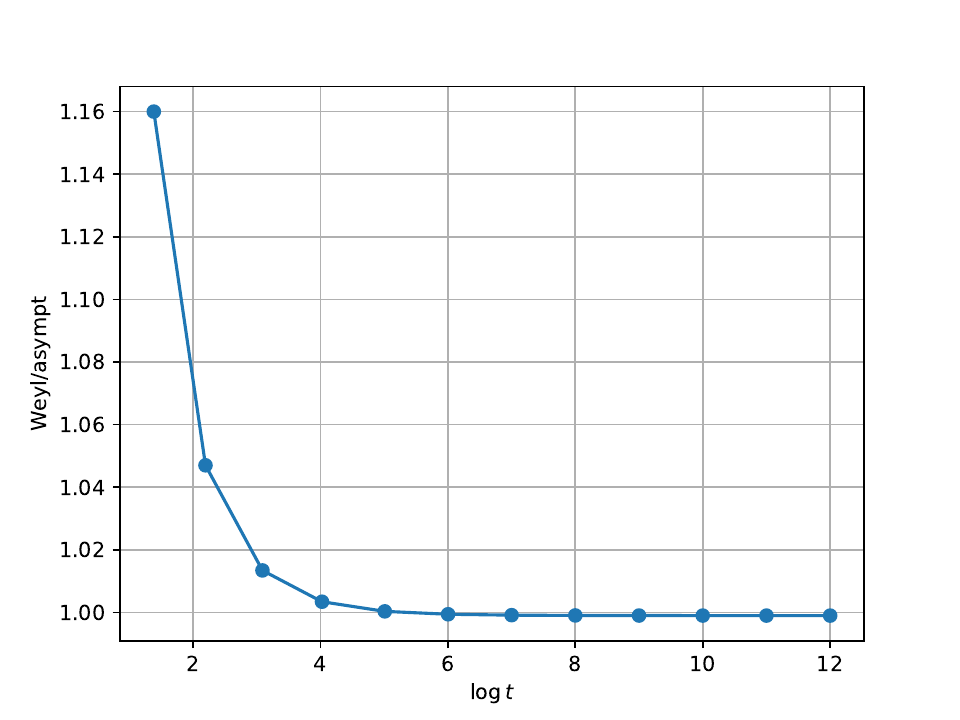}
	\caption{
		$\log\dim\lambda$ to $\frac12 N^2 \log t$ ratio for $n=1000000,\,K=1000,\,\delta=2$ (source: \cite{github}).
	}
	\label{Fig:dladder_n10E5}
\end{figure}

Here we present some numeric illustrations of the scaling formula \eqref{ladder_sca}. For the special cases of \textquote{double-ladder} diagrams with equal \(\delta=\tilde{\lambda}_{i-1}-\tilde{\lambda}_{i}\) for $i=2\ldots K$ we choose:
\begin{itemize}
	\item $n=400000,\,K=500,\,\delta=2$;
	\item $n=500000,\,K=500,\,\delta=4$;
	\item $n=1000000,\,K=1000,\,\delta=2$,	
\end{itemize}
and, making rescaling \eqref{rescaling}, numerically calculate the ratio \(\frac{\log\dim\lambda(t)}{\frac12 N^2\log t}\)
	of the direct application of the Weyl formula \eqref{Weyl_d} with its main asymptotic \eqref{ladder_sca}, as an example see  
	Fig.~\ref{Fig:dladder_n10E5}.

\section{More on the Goldstone bosons}

\subsection{Folding of Dynkin diagrams and breaking of $SU(N)$
	\label{app:goldstone}}

\begin{figure}[t]
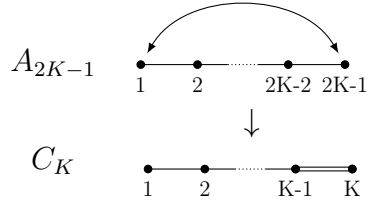

	\centering
	\begin{tabular}{ccc}
		$A_{2K-1}$&
		\begin{dynkinDiagram}[edge length=.75cm]{A}{**.**}
			\invol{1}{4}
			\dynkinLabelRoot{1}{1}
			\dynkinLabelRoot{2}{2}
			\dynkinLabelRoot{3}{$2K-2$}
			\dynkinLabelRoot{4}{$2K-1$}
		\end{dynkinDiagram}
		&\\
		&$\downarrow$&\\
		$C_K$&
		\begin{dynkinDiagram}[edge length=.75cm]{A}{**.**}
			\dynkinLabelRoot{1}{1}
			\dynkinLabelRoot{2}{2}
			\dynkinLabelRoot{3}{$K-1$}
			\dynkinLabelRoot{4}{$K$}
			\dynkin[at=(root 3),arrows=false]B2
		\end{dynkinDiagram}
		&
	\end{tabular}
	\caption{Folding of the Dynkin diagrams corresponding to the
		\(\mathfrak{sp}(K)\subset \mathfrak{sl}(2K)\) embedding.}
	\label{fi:foldingDy}
\end{figure}

Changing the basis by permutation of even vectors \(P=P_{2,N}P_{4,N-2}\ldots P_{N/2-2,N/2+2}\) one brings the VEV of $b$ in Eq.~\eqref{b_VEV} to the anti-diagonal form
\begin{equation}
	\Omega = PJP = \left(
	\begin{array}{cccc}
		0 &  \ldots & 0 & \epsilon \\
		0 &   \ldots & \epsilon & 0 \\
		\vdots &  \ddots  & & \vdots \\
		\epsilon & \ldots &  & 0 
	\end{array}
	\right)
\end{equation}
also satisfying  \(\Omega+\Omega^T=0\) and \(\Omega^2=-1\). In this basis the \(\mathfrak{sp}(K)\) subalgebra
\begin{equation}
	\label{XO}
	\Omega X+X^T\Omega=0,\ \ \ X\in \mathfrak{sl}(N)
\end{equation}
is given by \(A_{2K-1}\to C_K\) folding, 
i.e. is generated by invariant under involution \(\sigma (x_i)=\sigma(x_{2K-i})\) for the Cartan-Chevalley \(x=\{h,e,f\}\) generators combinations
\begin{equation}
	x_i+x_{2K-i},\ \ \ i=1,\ldots,K-1
\end{equation}
generating, together with \(x_K=\sigma(x_K)\) the \(\mathfrak{sp}(K)\) subalgebra of the matrices
\begin{equation}
	\label{Xsp}
	X = \left(
	\begin{array}{cc}
		A & B \\
		C & {}^tA
	\end{array}\right) \in \mathfrak{sp}(K)
\end{equation}
where \(K\times K\) blocks satisfy \({}^tB=B\) and \({}^tC=C\), and \({}^tA\) means transposition w.r.t. anti-diagonal. The dimension of \eqref{Xsp} is
exactly
\begin{equation}
	2\frac{K(K+1)}2 + K^2 = K(2K+1) = \frac{N(N+1)}{2} = \dim \mathfrak{sp}(K=N/2)
\end{equation}
where \(K^2\) comes from the \(A\in \mathfrak{gl}(K)\subset \mathfrak{sp}(K)\) subalgebra, corresponding to ``\(A_K\)-part'' of the \(C_K\) Dynkin diagram, see Fig.~\ref{fi:foldingDy}, while \(K(K+1)/2\) is the number of extra positive (or negative) roots, generated in block \(B\) by commutators with \(e_K\) (or with \(f_K\) in block \(C\)).

The coset space \(SU(N)/Sp(K)\) corresponds to the Goldstone modes, arising due to spontaneous symmetry breaking. Its dimension (together with extra broken \(U(1)_B\))  is
\begin{equation}
	\label{numGold2}
	N^2-\frac{N(N+1)}{2} = \frac{N(N-1)}{2}
\end{equation}
Indeed, the matrices from the coset space \(\mathfrak{sl}(N)/\mathfrak{sp}(K=N/2)\) satisfy the anti-involution constraint  \(\sigma (x_i)=-\sigma(x_{2K-i})\) and therefore are given by 
\begin{equation}
	\label{Xcoset}
	\tilde{X} = \left(
	\begin{array}{cc}
		\tilde{A} & \tilde{B} \\
		\tilde{C} & -{}^t\tilde{A}
	\end{array}\right) \in \mathfrak{sl}(N)/\mathfrak{sp}(K)
\end{equation}
with \({}^t\tilde{B}=-\tilde{B}\) and \({}^t\tilde{C}=-\tilde{C}\) and there is an extra condition for the matrix
\begin{equation}
	\tilde{A} = \left(
	\begin{array}{cccc}
		\ast &  \ldots & \ast &\ast \\
		\ast &   \ldots & \ast & \ast \\
		\vdots &  \ddots  & & \vdots \\
		\ast & \ldots &  & 0 
	\end{array}
	\right)
\end{equation}
coming from factoring out \(h_K\) by \(\sigma(h_K)=-h_K\). The dimension of the space \eqref{Xcoset} is (up to \(U(1)_B\)) the number of Goldstone modes \eqref{numGold2}
\begin{multline}
	2\frac{K(K-1)}2 + K^2-1 = 2K^2-K-1 =  \frac{N(N-1)}{2}-1 =
	\\
	=\dim \mathfrak{sl}(N)- \dim \mathfrak{sp}(K)=  \#_{\mathrm{Gold}}-1
	\label{gscount}
\end{multline}	

\subsection{Numeric check}
\label{app:num_goldstone}

Let us finally present the results of numeric verification of \eqref{gscount}. 
Consider \eqref{b_trans} in the infinitesimal form 
\begin{equation}
	b'=UbU^T\approx b+i\delta b=b+i\sum_{i=1}^{N^2-1}\alpha^a\{T^a b+b (T^a)^T\},
\end{equation}
where $\{T^a\}$ are generators of $\mathfrak{su}_N$. To count the number of Goldstone modes one has to find the number of linearly independent parameters, solving
\begin{equation}
	\delta b=\sum_{i=1}^{N^2-1}\{T^a b+b (T^a)^T\}\alpha^a=0,
	\label{del_b}
\end{equation}
which is a system of $N^2-1$ linear equations for $\{\alpha^a\}$. 

Calculating the rank of matrix in \eqref{del_b} gives the number of Goldstones. 
This algorithm was implemented in \textit{Mathematica} package \cite{github} and confirms \eqref{gscount} for every even $N$ (calculations for $N=\{2, 4, 6, 8, 10,12, 50,\ldots\}$ give the Goldstone mode's numbers $\#_{\mathrm{Gold}}-1=\{0, 5, 14, 27, 44, 65, 1224,\ldots\}$ correspondingly).



\begin{thebibliography}{99}
	
	\bibitem{HT1}
	A.~Hanany and D.~Tong,
	{\em Vortices, instantons and branes,}
	JHEP {\bf 0307}, 037 (2003).
	[hep-th/0306150].
	
	\bibitem{ABEKY}
	R.~Auzzi, S.~Bolognesi, J.~Evslin, K.~Konishi and A.~Yung,
	{\em Non-Abelian superconductors: Vortices and
		confinement in ${\mathcal N}=2$  SQCD,}
	Nucl.\ Phys.\ B {\bf 673}, 187 (2003).
	[hep-th/0307287].
	
	\bibitem{SYmon}
	M.~Shifman and A.~Yung,
	{\em Non-Abelian string junctions as confined monopoles,}
	Phys.\ Rev.\ D {\bf 70}, 045004 (2004)
	[hep-th/0403149].
	
	\bibitem{HT2}
	A. Hanany and D. Tong,
	{\em Vortex strings and four-dimensional gauge dynamics,}
	JHEP {\bf 0404}, 066 (2004)
	[hep-th/0403158].
	
	\bibitem{Trev}
	D.~Tong, {\em TASI Lectures on Solitons,}
	arXiv:hep-th/0509216.
	
	\bibitem{Jrev}
	M.~Eto, Y.~Isozumi, M.~Nitta, K.~Ohashi and N.~Sakai,
	{\em Solitons in the Higgs phase: The moduli matrix approach,}
	J.\ Phys.\ A  {\bf 39}, R315 (2006)
	[arXiv:hep-th/0602170].
	
	\bibitem{SYrev}
	M.~Shifman and A.~Yung,
	{\em Supersymmetric Solitons and How They Help Us Understand Non-Abelian Gauge Theories,}
	Rev.\ Mod.\ Phys.\  {\bf 79}, 1139 (2007)
	[hep-th/0703267]; for an expanded version see
	{\sl Supersymmetric Solitons,}
	(Cambridge University Press, 2009).
	
	\bibitem{Trev2}
	D.~Tong,
	{\em Quantum Vortex Strings: A Review,}
	Annals Phys.\  {\bf 324}, 30 (2009)
	[arXiv:0809.5060 [hep-th]].
	
	\bibitem{SYcstring} 
	M.~Shifman and A.~Yung,
	{\em Critical String from Non-Abelian Vortex in Four Dimensions,}
	Phys.\ Lett.\ B {\bf 750}, 416 (2015)
	[arXiv:1502.00683 [hep-th]].
	
	
	\bibitem{KSYconifold}
	P.~Koroteev, M.~Shifman and A.~Yung,
	{\em  Non-Abelian Vortex in
		Four Dimensions as a Critical  String on a Conifold},
	Phys.\ Rev.\ D {\bf 94} (2016) no.6,  065002
	[arXiv:1605.08433 [hep-th]].
	
	
	
	\bibitem{Candel}
	P.~Candelas and X.~C.~ de la Ossa,
	{\em Comments on conifolds,}
	Nucl. \ Phys. \ {\bf B342}, 246 (1990).
	
	\bibitem{NVafa}
	A.~Neitzke and  C.~Vafa,
	{\em Topological strings and their physical applications},
	arXiv:hep-th/0410178.
	
	\bibitem{SYlittles} 
	M.~Shifman and A.~Yung,
	{\em Critical Non-Abelian Vortex in Four Dimensions and Little String Theory,}
	Phys.\ Rev.\ D {\bf 96}, no. 4, 046009 (2017)
	[arXiv:1704.00825 [hep-th]].
	
	
	\bibitem{Ivanov}
	E.~ Ivanov and S.~Krivonos, 
	{\em U(1) supersymmetric extension of the Liouville equation,}
	Lett. \ Math. \ Phys.\ {\bf  7},  523 (1983). 
	
	\bibitem{KutSeib}
	D.~Kutasov and N.~Seiberg,
	{\em Noncritical Superstrings,}
	Phys.\ Lett.\ B {\bf 251}, 67 (1990).
	
	\bibitem{Kutasov}
	D.~Kutasov,
	{\em Introduction to Little String Theory}, published in
	{\sl Superstrings and Related Matters 2001},  Proc. of the ICTP Spring School 
	of Physics, Eds. C. Bachas, K.S. Narain, and   S. Randjbar-Daemi, 2002,  pp.165-209.
	
	\bibitem{GVafa}
	D.~Ghoshal and  C.~Vafa, 
	{\em c = 1 String as the Topological Theory of the Conifold},
	Nucl.\ Phys.\ B {\bf 453}, 121 (1995)
	[hep-th/9506122].
	
	\bibitem{GivKut}
	A.~Giveon and D.~Kutasov,
	{\em Little String Theory in a Double Scaling Limit},
	JHEP {\bf 9910}, 034 (1999)
	[hep-th/9909110].
	
	\bibitem{GivKutP}
	A.~Giveon, D.~Kutasov and O.~Pelc,
	{\em Holography for Noncritical Superstrings},
	JHEP {\bf 9910}, 035 (1999)
	[hep-th/9907178].
	
	\bibitem{SYlittmult}  
	M.~Shifman and A.~Yung,
	{\em Hadrons of $\mathcal N=2$ Supersymmetric QCD in Four Dimensions from Little String Theory,}
	Phys.\ Rev.\ D {\bf 98}, no. 8, 085013 (2018)
	[arXiv:1805.10989 [hep-th]].
	
	\bibitem{GIMMY}
	P.~Gavrylenko, E.~Ievlev, A.~Marshakov, I.~Monastyrskii and   A.~Yung,
	{\em 2D Sigma Models on Non-compact Calabi-Yau and $N=2$ Liouville Theory,}
	Phys.\ Rev.\ D {\bf 111}, 106003 (2025)
	arXiv:2307.02929[hep-th]
	
	\bibitem{Y_mass_Liouville}
	A.~Yung,
	{\em Flowing between string vacua for the critical non-Abelian vortex
		with a deformation of N = 2 Liouville theory,}
	Phys. \ Rev. \ D {\bf 110},  025004 (2024)
	[arXiv:2403.20099[hep-th]]
	
	\bibitem{IMSY}
	E.~Ievlev, A.~Marshakov, G.~Sumbatian and A.~Yung,
	{\em Critical non-Abelian vortex string and 2D N=2 black hole,}
	Phys. Rev. D \textbf{112} (2025) no.10, 105010
	doi:10.1103/rz16-99g6
	[arXiv:2508.12972 [hep-th]].
	
	\bibitem{Giveon}
	A.~Giveon, 
	{\em Target space duality and stringy black holes,}
	Mod.\ Phys.\ Lett. \textbf{A6},  2843 (1991).
	
	\bibitem{DijVerVer}
	R.~Dijkgraaf, H.~Verlinde and E.~Verlinde,
	{\em String propagation in a black hole geometry,}
	Nucl. \ Phys. \ {\bf B 371} ,  269 (1992).
	
	
	\bibitem{Wbh}
	E.~Witten,
	{\em String Theory and Black Holes,}
	Phys.\ Rev.\ D {\bf 44}, 314 (1991).
	
	
	
	\bibitem{MukVafa}
	S.~Mukhi and C.~Vafa,
	{\em Two-dimensional black hole as a topological coset model of c = 1 string theory},
	Nucl. \ Phys.\ {\bf B407}  667, (1993)
	[arXiv: hep-th/9301083].
	
	
	\bibitem{OoguriVafa95}
	H.~Ooguri and C.~Vafa,
	{\em Two-Dimensional Black Hole and Singularities of CY Manifolds,}
	Nucl.\ Phys.\ B {\bf 463}, 55 (1996)
	[hep-th/9511164].
	
	\bibitem{HoriKapustin}
	K.~Hori and A.~Kapustin,
	{\em Duality of the fermionic 2-D black hole and N=2 Liouville theory as mirror symmetry,}
	JHEP {\bf 0108}, 045 (2001)
	[hep-th/0104202].
	
	\bibitem{SW2} 
	N.~Seiberg and E.~Witten,
	{\em Monopoles, duality and chiral symmetry breaking in N=2 supersymmetric QCD,}
	Nucl. Phys. {\bf B431}, 484  (1994)
	[hep-th/9408099].
	
	
	\bibitem{FradShen}
	E.~H.~Fradkin and S.~H.~Shenker, 
	{\em Phase Diagrams of Lattice Gauge Theories with
		Higgs Fields,}
	Phys. \ Rev. \ D {\bf 19}, 3682 (1979).
	
	\bibitem{FI}
	P.~Fayet and J.~Iliopoulos,
	``Spontaneously Broken Supergauge Symmetries and Goldstone Spinors,''
	Phys.\ Lett.\  B {\bf 51}, 461 (1974).
	
	\bibitem{SYi_of_c}
	M.~Shifman and A.~Yung, 
	{\em Lessons from supersymmetry:
		“Instead-of-Confinement” mechanism,}
	Int. \ J. \ Mod.\  Phys. {\bf A29}, 1430064 (2014), arXiv:1410.2900 [hep-th].
	
	
	\bibitem{ISY_b_baryon} 
	E.~Ievlev, M.~Shifman and A.~Yung,
	{\em String baryon in four-dimensional
		N=2 supersymmetric QCD from the 2D-4D correspondence,}
	Phys.\ Rev.\ D {\bf 102}, 054026 (2020)
	[arXiv:2006.12054 [hep-th]].
	
	\bibitem{W93}
	E.~Witten,
	{\em Phases of N = 2 theories in two dimensions,}
	Nucl.\ Phys.\ B {\bf 403}, 159 (1993).
	[hep-th/9301042].
	
	\bibitem{AchVas}
	For a review see e.g. A.~Achucarro and T.~Vachaspati,
	{\em Semilocal and electroweak strings,}
	Phys.\ Rept.\  {\bf 327}, 347 (2000)
	[hep-ph/9904229].
	
	\bibitem{SYsem}
	M.~Shifman and A.~Yung,
	{\em Non-Abelian semilocal strings in  ${\mathcal N} = 2$ supersymmetric QCD,}
	Phys.\ Rev.\  D {\bf 73}, 125012 (2006)
	[arXiv:hep-th/0603134].
	
	\bibitem{Jsem}
	M.~Eto, J.~Evslin, K.~Konishi, G.~Marmorini, et al.,
	{\em On the moduli space of semilocal strings and lumps,}
	Phys.\ Rev.\  D {\bf 76}, 105002 (2007)
	[arXiv:0704.2218 [hep-th]].
	
	
	\bibitem{SVY}
	M.~Shifman, W.~Vinci and A.~Yung, 
	{\em Effective World-Sheet Theory for Non-Abelian Semilocal 
		Strings in ${\mathcal N} = 2$ Supersymmetric QCD,}
	Phys.\ Rev.\ D {\bf 83}, 125017 (2011)
	[arXiv:1104.2077 [hep-th]].
	
	\bibitem{Dorey}
	N.~Dorey,
	{\em The BPS spectra of two-dimensional supersymmetric gauge theories with  twisted mass terms,}
	JHEP {\bf 9811}, 005 (1998) [hep-th/9806056].
	
	\bibitem{DoHoTo}
	N.~Dorey, T.~J.~Hollowood and D.~Tong,
	{\em The BPS spectra of gauge theories in two and four dimensions,}
	JHEP {\bf 9905}, 006 (1999)
	[arXiv:hep-th/9902134].
	
	
	\bibitem{IYcorrelators}
	E.~Ievlev and A.~Yung,
	{\em Critical Non-Abelian vortex and holography for little string theory,}
	Phys. Rev. D \textbf{104},  114033 (2021)
	[arXiv:2110.08546 [hep-th]].	
	
	\bibitem{Nakayama}
	Y.~Nakayama,
	{\em Liouville field theory: A Decade after the revolution,}
	Int. J. Mod. Phys. A \textbf{19}, 2771-2930 (2004)
	[arXiv:hep-th/0402009 [hep-th]].
	
	\bibitem{Teschner:1999ug}
	J.~Teschner,
	{\em Operator product expansion and factorization in the $H_3^+$ WZNW model,}
	Nucl. Phys. B \textbf{571}, 555-582 (2000)
	[arXiv:hep-th/9906215 [hep-th]].
	
	\bibitem{LSZinLST}
	O.~Aharony, A.~Giveon and D.~Kutasov,
	{\em LSZ in LST,}
	Nucl. Phys. B \textbf{691}, 3-78 (2004)
	[arXiv:hep-th/0404016 [hep-th]].
	
	\bibitem{DixonPeskinLy}
	L.~J.~Dixon, M.~E.~Peskin and J.~D.~Lykken,
	{\em N=2 Superconformal Symmetry and SO(2,1) Current Algebra,}
	Nucl.\ Phys.\ B {\bf 325}, 329 (1989).
	
	\bibitem{Petrop}
	P.M.S.~Petropoulos,
	{\em Comments on SU(1,1) string theory},
	Phys. \ Lett.\ {\bf B236}, 151 (1990).
	
	\bibitem{Hwang}
	S.~Hwang,
	{\em Cosets as Gauge Slices in SU(1,1) Strings,}
	Phys. \  Lett.\ {\bf B276} 451, (1992) 
	[arXiv:hep-th/9110039].
	
	\bibitem{EGPerry}
	J.~M.~Evans, M.~R.~Gaberdiel and M.~J.~Perry,
	{\em The no ghost theorem for AdS(3) and the stringy exclusion principle,}
	Nucl.\ Phys.\ B {\bf 535}, 152 (1998)
	[hep-th/9806024].
	
	
	\bibitem{EGPerry-rev}
	J.M.~Evans, M.R.~Gaberdiel and M.J~Perry,
	{\em The no-ghost theorem and strings on AdS$_3$},
	[hep-th/9812252], published in Proc. 1998 ICTP Spring School of Physics
	{\sl Nonperturbative Aspects of Strings, Branes and Supersymmetry},  Eds. M. J. Duff {\em et al.},
	pp. 435-444.
	
	\bibitem{Strominger_95}
	A. Strominger,
	{\em Massless black holes and conifolds in string theory,}
	Nucl. \ Phys. \ B {\bf 451}, 96 (1995)
	hep-th/9504090.  
	
	\bibitem{W79} 
	E.~Witten,
	{\em Instantons, The Quark Model, And The 1/N Expansion,}
	Nucl.\ Phys.\ B {\bf 149}, 285 (1979).
	
	\bibitem{Mertens}
	T.~G.~Mertens, 
	{\em Hagedorn String Thermodynamics in Curved Spacetimes and near Black
		Hole Horizons,}
	Ph.D. thesis, Gent U. (2015), arXiv:1506.07798 [hep-th]

\bibitem{GreenSchWitt}
M.~ B.~ Green, J.~ H.~ Schwarz, and E.~ Witten, 
{\em Superstring Theory,}
 Cambridge University Press, Cambridge, England,
(1987).
	
	
	\bibitem{Hagedorn}
	R.~Hagedorn, {\em Statistical thermodynamics of strong interactions at high-energies,}  Nuovo Cim.
	Suppl. {\bf 3}, 147 (1965).
	
	
	\bibitem{FM}
	V.~Fainberg and A.~Marshakov, {\em A propagator for the fermionic string}, 
	Phys.\ Lett.\ {\bf B 211} (1988) 81
	
	\bibitem{Susskind}
	L.~Susskind, 
	{\em Some speculations about black hole entropy in string theory,} 
	arXiv:hep-th/9309145.
	
	
	
	\bibitem{HorowPolch}
	G.~T.~Horowitz and J.~Polchinski, 
	{\em A correspondence principle for black holes and strings,}
	Phys. \ Rev. \ D {\bf 55}, 6189 (1997) 
	[arXiv:hep-th/9612146]
	
	\bibitem{GivKutRabin}
	A.~Giveon, D.~Kutasov, E.~Rabinovici and  A.~Sever, 
	{\em Phases of quantum gravity in AdS$_3$ and linear dilaton backgrounds,}
	Nucl. \ Phys. \ {\bf B719}, 3 (2005) 
	[arXiv:hep-th/0503121]
	
	\bibitem{AtickWitt}
	J.~J.~ Atick and E. ~Witten,
	{ \em The Hagedorn Transition and the Number of Degrees of
		Freedom of String Theory,}
	Nucl. \ Phys. \   {\bf B 310},  291 (1988)
	
	\bibitem{Kut_wind_condens}
	D.~Kutasov, 
	{\em\ Accelerating branes and the string/black hole transition,}
	arXiv:hep-th/0509170
	
	\bibitem{KazakKostovKut}
	V.~Kazakov, I.~K.~Kostov, and D.~Kutasov, 
	{\em A Matrix model for the two-dimensional black hole,}
	Nucl.  \ Phys. \ {\bf B 622}, 141 (2002), arXiv:hep-th/0101011
	
	\bibitem{Malda_BH_hagedorn}
	Y. Chen and J. Maldacena, 
	{\em String scale black holes at large D, }
	JHEP \ {\bf 01},  095 (2022)
	[arXiv:[2106.02169 [hep-th]].
	
	\bibitem{partitions}
	George E.~Andrews,
	{\em The Theory of Partitions,}
	Cambridge University Press, 1976
	
	\bibitem{github}
	G.~Sumbatian, \texttt{uN\_from\_BH}
	\url{https://github.com/Defa777/uN_from_BH}
	
	
\end{thebibliography}
\end{document}